# Modular Reactor for *In Situ* X-ray Scattering, Spectroscopy, and ATR-IR Studies of Solvothermal Nanoparticle Synthesis

**Sani Y. Harouna-Mayer,[a,b,+] Melike Gumus Akcaalan,[a,+] Jagadesh Kopula Kesavan,[a,b] Tjark R. L. Groene,[a] Lars Klemeyer,[a,b] Sarah-Alexandra Hussak,[a,c] Lukas Grote,[a] Davide Derelli,[a] Francesco Caddeo,[a] Cecilia Zito,[a,b] Paul Stützle,[a] Dorota Speer,[a] Ann-Christin Dippel,[c] Blanka Detlefs,[d] Yannik Appiarius,[e] Axel Jacobi von Wangelin,[e] Dorota Koziej,[a,b,*]**

[+] Those authors contributed equally.

[a] University of Hamburg, Institute for Nanostructure and Solid-State Physics, Center for Hybrid Nanostructures (ChyN), Luruper Chaussee 149, 22761 Hamburg, Germany

[b] The Hamburg Center for Ultrafast Imaging, 22761 Hamburg, Germany

[c] Deutsches Elektronen Synchrotron DESY, Notkestraße 85, 22607 Hamburg Germany

[d] ESRF, The ESRF, The European Synchrotron, 71 Avenue des Martyrs, CS40220, 38043 Grenoble, Cedex 9, France

[e] University of Hamburg, Department of Chemistry, Martin Luther King Platz 6, 20146 Hamburg, Germany

* Corresponding author. Mail: dorota.koziej@uni-hamburg.de

**Synopsis** A reactor is described, whose modular design enables *in situ* x-ray scattering and spectroscopy, infrared spectroscopy, gas and liquid injection under autoclave-like conditions and magnetic stirring for solvothermal synthesis up to 200 °C and 8 bar.

**Abstract** Understanding the chemical processes that occur during the solvothermal synthesis of functional nanomaterials is essential for their rational design and optimization for specific applications. However, these processes remain poorly understood, primarily due to the limitations of conventional *ex situ* characterization techniques and the technical challenges associated with *in situ* studies, particularly the design and implementation of suitable reactors. Here, we present a versatile cell suitable for *in situ* x-ray scattering, x-ray spectroscopy and infrared spectroscopy studies performed during solvothermal synthesis under autoclave-like, inert conditions. The reactor enables precise control of the temperature between -20 °C and 200 °C, pressures up to 8 bar, magnetic stirring, and injection of gas or liquids. The reactor's capabilities are demonstrated by comprehensively studying the solvothermal synthesis of magnetite nanoparticles from iron acetylacetonate in benzyl alcohol through *in situ* x-ray scattering and spectroscopy, and ATR-IR spectroscopy.

**Keywords: reactor design; Solvothermal synthesis; Complementary *in situ* x-ray absorption spectroscopy and x-ray scattering; Multimodal *in situ* infrared spectroscopy**

## 1. Introduction

Solvothermal synthesis is a versatile, fast, energy-efficient and scalable method for producing a wide range of materials, in which chemical reactions between precursor(s) and solvent take place in a closed reaction vessel at elevated temperatures and pressures. Although it is widely used for synthesizing materials or composites e.g. for electronic devices and energy conversion, the underlying reaction mechanisms that govern key properties such as crystallinity, particle size and morphology are often poorly understood. (Heiligtag and Niederberger, 2013; Yin and Talapin, 2013; Boles *et al.*, 2016) In conventional laboratory experiments, the understanding of these mechanisms is typically limited to *ex situ* analysis of the final products or quenching the reaction at different reaction times. However, a deeper insight into these processes is crucial for the rational design of materials with tailor-made properties. (Cansell and Aymonier, 2009; Deshmukh and Niederberger, 2017; Terraschke, 2023)

For this reason, solvothermal synthesis can be studied *in situ*, by using high flux x-ray radiation from synchrotron light sources, allowing real-time monitoring of the reaction and providing a comprehensive insight into the chemical steps involved. (Jensen *et al.*, 2014; D. Bøjesen and B. Iversen, 2016) This approach requires specialized reaction vessels equipped with heating capabilities and additional features tailored to the specific reaction and technique being used. X-ray methods are particularly effective for *in situ* investigation of solvothermal synthesis as they provide valuable information about nucleation, growth and formation of intermediate states. For example, x-ray absorption spectroscopies (XAS) probe the electronic structure and the chemical environment of the absorbing atom. (Koziej, 2016) Wide-angle x-ray scattering (WAXS) methods, such as powder x-ray diffraction (PXRD) and total scattering (TS), provide insights into the atomic structure while small angle x-ray scattering (SAXS) enables to study particle size, polydispersity, morphology, and assembly. (Nyborg Broge *et al.*, 2020; Zhu *et al.*, 2021) A single technique can mostly not capture the full complexity of the reaction mechanisms across multiple length and time scales. It is rather the simultaneous application of different techniques in a multimodal or complementary approach that enables a more comprehensive understanding - e.g. by linking electronic, atomic, and morphological evolution throughout the reaction. (Grosso *et al.*, 2004; Andersen *et al.*, 2017; Grote *et al.*, 2021) However, each x-ray technique imposes distinct and often conflicting demands on the reactor design and specifications.

In order to meet these demands, several key challenges must be addressed when designing a tailor-made reactor for *in situ* x-ray studies:

(*i*) Window material. The material must be x-ray transparent at the required energies, chemically inert, and stable at elevated temperatures and pressures. It should also exhibit a low or easily subtractable background signal. Various materials are used depending on the technique and required reaction conditions:

- Polyimide is flexible, highly x-ray transparent and chemically inert, making it a common choice for both XAS and scattering experiments conducted under moderate conditions. (Beyer *et al.*, 2014; Staniuk *et al.*, 2014; Thum *et al.*, 2024)
- Polyether ether ketone (PEEK) offers excellent x-ray transparency along with high chemical and mechanical stability and is thus suitable for use at elevated temperatures and pressures. While it performs well in XAS setups, its semicrystalline structure produces strong background signals, making it less suitable for scattering experiments. (Grunwaldt *et al.*, 2005; Grote *et al.*, 2021; Klemeyer *et al.*, 2024)
- Fused silica is also chemically inert and can withstand high temperatures and pressures but has a slightly lower x-ray transparency than PEEK. Though, it has a clean background signal for accurate background subtraction in scattering data. It is the commonly preferred window material for PXRD and TS experiments. (Şahin *et al.*, 2021; Roelsgaard *et al.*, 2023; Derelli *et al.*, 2024)
- Sapphire is suitable for the extreme temperatures and pressures of PXRD experiments. Its single crystal reflections can be avoided by masking the detector or angular adjustments. However, sapphire is not favorable for TS experiments due to its strong diffuse scattering signal. (Chupas *et al.*, 2008; Becker *et al.*, 2010; Roelsgaard *et al.*, 2023)
- Beryllium is used in some experimental setups for its exceptional x-ray transparency and mechanical stability. However, its high toxicity necessitates strict safety protocols. (Grunwaldt *et al.*, 2005; Bare *et al.*, 2007; Testemale and Brugger, 2024)

(*ii*) Geometry of the reactor and (*iii*) precise control over reaction parameters such as temperature, pressure, and (possibly) injection of reagents or gases to the chemical reactions inside the reactor. It is highly desirable to mimic the laboratory setups of preparative-scale chemical reactions, as variations in temperature or pressure profiles and sample volume can significantly influence the reaction mechanism and outcome. At the same time, the reactor must meet the space constraints of the beamline to allow *in situ* analytical experiments. Most *in situ* setups for x-ray scattering use capillaries, where small sample volumes are heated with hot air blowers or resistive heaters. (Chupas *et al.*, 2008; Becker *et al.*, 2010; Roelsgaard *et al.*, 2023) This approach requires little space and allows rapid heating and good control over the reaction parameters. It can also be extended with gas or liquid injection devices. However, such setup usually does not allow stirring of the reaction solution and the small volumes are usually not sufficient for post-synthesis analysis. Other setups use large reaction reservoirs from which small aliquots of the solution are circulated through a capillary for measurement, thereby mitigating beam damage. (Yi *et al.*, 2015; Şahin *et al.*, 2021) Bulky setups adapting commercial microwave reactors for the technical requirements of *in situ* measurements have also been reported. (Tominaka *et al.*, 2018; Yamada *et al.*, 2019)

(*iv*) Safety of the beamline environment, especially when working at extreme conditions or with hazardous or toxic materials such as Be. (Testemale and Brugger, 2024)

Table 1 summarizes selected literature reports on reactors enabling *in situ* analysis by XAS, WAXS, and SAXS.

In an effort to address the aforementiofned diverse requirements of a modular and versatile reactor, we developed a tailor-made setup for complementary experiments at synchrotron facilities. The reactor is compatible with a wide range of x-ray techniques, including XAS, PXRD, TS, and SAXS, and supports solvothermal synthesis under inert, autoclave-like conditions. By exchanging the reactor inlet, the setup can be tailored for XAS experiments using a PEEK inlet or for scattering using glass inlets with various wall thicknesses – balancing background contribution and pressure resistance up to 8 bar. The system allows magnetic stirring and precise control of the heating ramp rate and reaction temperatures up to 200 °C or cooling to –20 °C via Peltier elements. Additional inlet modifications also support gas or liquid injection and integration of a fiber optic attenuated total reflection (ATR) probe for simultaneous ATR infrared spectroscopy (ATR-IR) measurements. The reactor achieves sub-minute time resolution for *in situ* XAS and ATR-IR, and one-second resolution for x-ray scattering experiments.

This reactor has already been employed in several studies investigating the solvothermal synthesis of alloy (Derelli *et al.*, 2024), metal oxide (Grote *et al.*, 2021), metal sulfide (Klemeyer *et al.*, 2024), and metal nitride (Harouna-Mayer *et al.*, 2025) nanoparticles. In these studies, we leveraged the reactor's modular design and versatility across different x-ray techniques – exploiting, for instance, its low-background configuration for precise tracking of precursor conversion (Derelli *et al.*, 2024), its autoclave-like properties for spatially resolved synthesis at liquid–liquid interfaces at elevated pressure and temperature (Klemeyer *et al.*, 2025), and its air-tight conditions for sensitive materials (Harouna-Mayer *et al.*, 2025). Here, we further demonstrate the reactor's versatility by investigating as model system the partial reduction reaction of iron(III) acetylacetonate to $Fe_3O_4$ magnetite nanoparticles in benzyl alcohol at 180 °C (Pinna *et al.*, 2005), using *in situ* PXRD, TS, XAS, and ATR-IR. Moreover, we also added cooling and injection capability to the cell, broadening its application range.

**Table 1** Overview of reactors for *in situ* XAS, WAXS and SAXS measurements. The sample volume is categorized as *small* (e.g. capillary-based setups), *medium* (e.g. reactors with inlets), or *large* (e.g. flow reactors with external reservoirs). The classification reflects the typical sample quantity accessible for reaction monitoring or postmortem analysis. The "+" in the temperature column indicates the temperature employed in the respective study, rather than the maximum operational temperature, which was not specified in the respective report.

| Reference | Methods | $T_{max}$ (°C) | $p_{max}$ (bar) | Window material | Wall thickness (mm) | Sample volume | Stirring? |
|---|---|---|---|---|---|---|---|
| (Evans *et al.*, 1995) | WAXS | 230 | 28 | Steel | 2.0 | Medium | Yes |
| (Grunwaldt *et al.*, 2005) | XAS | 200 | 250 | PEEK, Be | 1.0, 0.5 | Medium | Yes |
| (Bare *et al.*, 2007) | XAS | 600 | 14 | Be | 0.75 | Small | No |
| (Chupas *et al.*, 2008) | WAXS, SAXS | 1000 | 138 | $Al_2O_3$, $SiO_2$, Polyimide | - | Small | No |
| (Fingland, Ribeiro and Miller, 2009) | XAS | 400<br>300 | 3.5<br>40 | Polyimide<br>$SiO_2$ | 0.3 – 1.2 | | No |
| (Becker *et al.*, 2010) | WAXS, SAXS | 450 | 400 | $Al_2O_3$<br>$SiO_2$ | 0.06 – 0.8 | Small | No |
| (van Beek *et al.*, 2011) | WAXS | 450+ | 20 | $SiO_2$ | 0.01 | Small | No |
| (Borkiewicz *et al.*, 2012) | WAXS, SAXS, XAS | 48 | - | Glassy carbon | - | | No |
| (Staniuk *et al.*, 2014) | WAXS, XAS | 180+ | - | PEEK | 0.5 | Medium | No |
| (Yi *et al.*, 2015) | SAXS | - | - | $SiO_2$ | 0.01 | Large | Yes |
| (Heidenreich *et al.*, 2017) | WAXS, XAS | 180 | - | $SiO_2$, Al | 1.0 – 1.5, 0.1 | Medium | No |
| (Hoffman *et al.*, 2018) | WAXS | 1000 | 35 | $SiO_2$ | >0.1 mm | Small | No |
| (Kalantzopoulos *et al.*, 2018) | WAXS | 700+ | - | $SiO_2$ | - | | No |
| | WAXS | 160+ | - | $SiO_2$ | - | Medium | Yes |
| (Şahin *et al.*, 2021) | WAXS | 700+ | - | $SiO_2$ | 0.05 | Large | Yes |
| (Wang *et al.*, 2022) | WAXS, SAXS, XAS | 327 | 40 | $SiO_2$ | 0.2 | Small | No |
| (Prinz *et al.*, 2023) | WAXS | 500+ | - | $SiO_2$ | 0.02 | Small | No |
| (Testemale *et al.*, 2024) | XAS | 1200 | 2000 | Glassy carbon, Be | - | Large | No |
| This work | WAXS, SAXS, XAS, ATR-IR | 200 | 8 | $SiO_2$, PEEK | 0.2 – 0.5 | Medium | Yes |

## 2. Reactor design

The reactor setup features a compact and user-friendly design optimized for safe operation. The assembled cross-sectional illustration of the reactor, including all dimensions, is shown in Figure 1(*a*). For a comprehensive understanding of the system's operating principle, it is beneficial to focus on the main elements individually. The detailed representation of the assembly is given in Figure 1(*b*). The reactor can be divided into three main components, *i.e.* the housing, the inlet, and the insulators.

The back and front center pieces were carved out from the aluminium housing in a conical shape to achieve a 45° opening angle enabling data acquisition in transmission geometry – *e.g.,* for high q total scattering (TS) data – or in reflection geometry, *e.g.,* high-energy resolution fluorescence detected XAS (HERFD-XAS). Heating elements are vital for the thermal performance of the reactor setup, as they influence the heating of the aluminium housing, the inlet, and the reaction solution. These factors determine the heating rate, final temperature, and temperature stability. They surround the inlet and are driven with a 24V-50W power supply (EA-PS 5040, Elektro-Automatik) in parallel connection. The generated dielectric heating is distributed across the area that surrounds the inlet, evenly via the thermal conductivity of the aluminium. The temperature is measured by a PT1000 model sensor (Honeywell) positioned close to the inlet and connected to an LS335 temperature controller. The measured temperature is instantly read from the temperature controller to a computer and monitored via a Python script.

The inlet works as a custom-made container for the reaction solution accommodation. It is used with a cap and a perfluoro elastomer (FFKM) O-ring to provide an air-tight environment. The inlet design offers two material options: polyether ether ketone (PEEK, Bieglo) and glass. The material is selected according to the measurement technique, and the x-ray window wall thickness can be adapted to the reaction requirements. PEEK is a favorable material for spectroscopy measurements due to its high x-ray transparency. Glass material is preferred for scattering experiments although its use presents challenges regarding mechanical stability. In careful tests, we observed that determination of the optimum glass thickness as a balance between satisfactory data collection and structural integrity at high-pressure conditions is quite challenging. To this end, we designed glass and PEEK inlets with 0.2, 0.3, and 0.5 mm wall thicknesses. The results demonstrate that 0.2 mm thickness of the PEEK inlet and 0.3 mm thickness of the glass inlet gave the most satisfactory data for the spectroscopy and scattering experiments at relatively high pressures, respectively. Figure 1(*c*) shows the various inlet and cap configurations for specific x-ray techniques and experimental requirements, with total and scan zone volumes, which are as follows:

(*i*) Glass inlet for scattering experiments under air-tight, autoclave-like conditions at elevated temperatures and pressures with a wall thickness of 0.3 mm.

(*ii*) Thin-walled (0.05 mm) glass vial embedded in a PEEK body for weak background scattering contributions.

(*iii*) PEEK inlet for XAS measurements with a wall thickness of 0.2 mm.

(*iv*) A modified hollow-screwed PEEK cap that enables the pressure sensor to measure pressure from a close position to the solution.

(*v*) A modified hollow Brass cap allowing the injection of reagents during the experiment through a septum.

(*vi*) A modified hollow Brass cap for a fiber-optical ATR-IR probe implementation, enabling multimodal ATR-IR measurements.

Further details about the reactor tailored for cooling, photographs of the caps designated for pressure and IR measurements is given in the supporting information.

The insulation and the sealing elements complete the air-tight environment, working as the supportive components. A tightly fastened aluminium component, mounted on top of the aluminium housing, assures complete sealing of the reactor. Top, bottom, and side PEEK insulators enhance the heat retention within the system by minimizing thermal exchange between the aluminium and surrounding air as much as possible. Properties such as high thermal stability, low thermal expansion coefficient, dimensional stability, and low thermal conductivity, which make PEEK a perfect thermal insulator, support all the thermal requirements of the closed system.

To ensure a homogeneous colloidal dispersion, a micro stirrer (Variomag Thermo Scientific™), is installed on the aluminium bottom plate directly beneath the aluminium housing. A micromagnetic stirring bar is located within the inlet, providing homogeneous mixing of the solution during the *in situ* measurements. This ensures a constant concentration of the scatterers across the cell and high data quality.

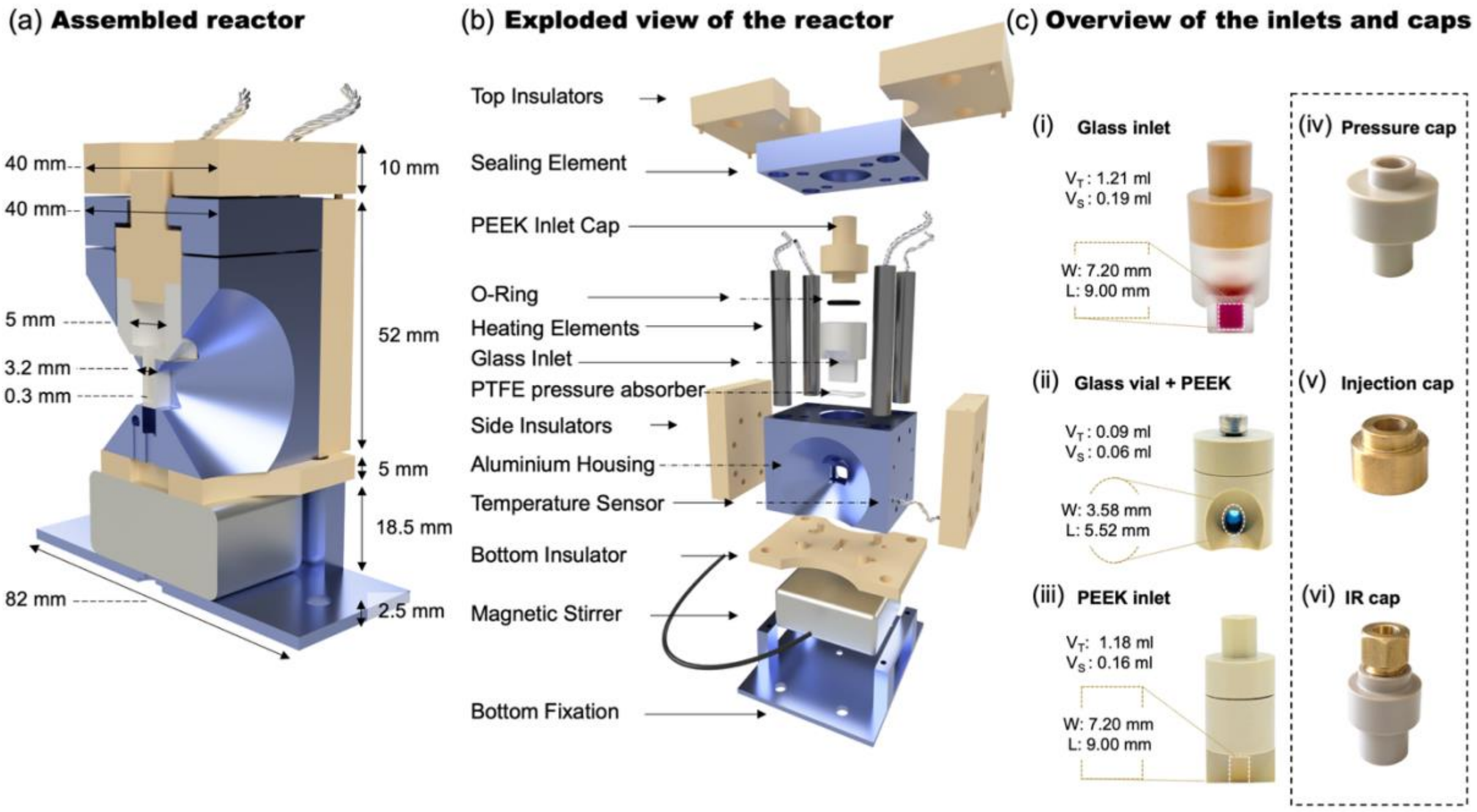

**Figure 1** Design of the reactor. (*a*) Cross-sectional rendering of the assembled reactor. (*b*) Exploded schematic showing individual components. (*c*) Photograph of the interchangeable inlet modules for different experimental configurations, $V_T$ and $V_s$ indicate total volume and scanning area volume, respectively. W and L state the width and length of the scanning area.

### 3. Performance of the reactor

We demonstrate the performance and versatility of the reactor through a series of characterization and reaction studies. First, we evaluate its thermal response, pressure stability, and x-ray transparency (Figure 2). We then estimate the signal-to-noise ratio of PXRD pattern of nanoparticle dispersions at different particle concentrations (Figure 3). Finally, we investigate the model system, the solvothermal synthesis of magnetite nanoparticles, in detail using a combination of complementary *in situ* techniques: PXRD and PDF analysis of TS data (Figure 4), ATR-IR (Figure 5), and HERFD-XAS (Figure 6). By combining all analytical data from these methods, we propose a reaction mechanism of the nanoparticle formation and postulate intermediate states during the transformation of iron(III) acetylacetonate, $Fe(acac)_3$, to magnetite, $Fe_3O_4$, in benzyl alcohol (Pinna *et al.*, 2005). These results highlight the reactor's capability to integrate multiple methods, providing a comprehensive understanding of complex reaction pathways.

Figure 2(*b*) shows photographs of the glass inlet during the $Fe_3O_4$ nanoparticle synthesis. The color change of the reaction solution from reddish $Fe(Acac)_3$ in benzyl alcohol (BnOH) to black

$Fe_3O_4$ is clearly visible during the heating ramp to 180 °C. Figure 2(*c*) depicts an IR image of the reactor window, illustrating uniform heat distribution in the BnOH-filled inlet at 180 °C. We further measure the temperature and pressure profiles during heating: the BnOH-filled inlet reaches 3.8 bar at 180 °C, while water-filled inlet reaches 7.9 bar at 150 °C – both without breaking the glass inlet with a 0.3 mm wall thickness, as shown in Figure 2(*d*). To assess the suitability of the glass inlet for x-ray experiments, we calculate its x-ray transmission both empty and filled with the common solvents water, ethanol, and BnOH. The results indicate reasonable transmission for x-ray energies above 20 keV, as shown in Figure 2(e). For comparison, Figure S4 shows the x-ray transmission of glass and PEEK inlets with different wall thicknesses. Details on the transmission calculations are given in the experimental section.

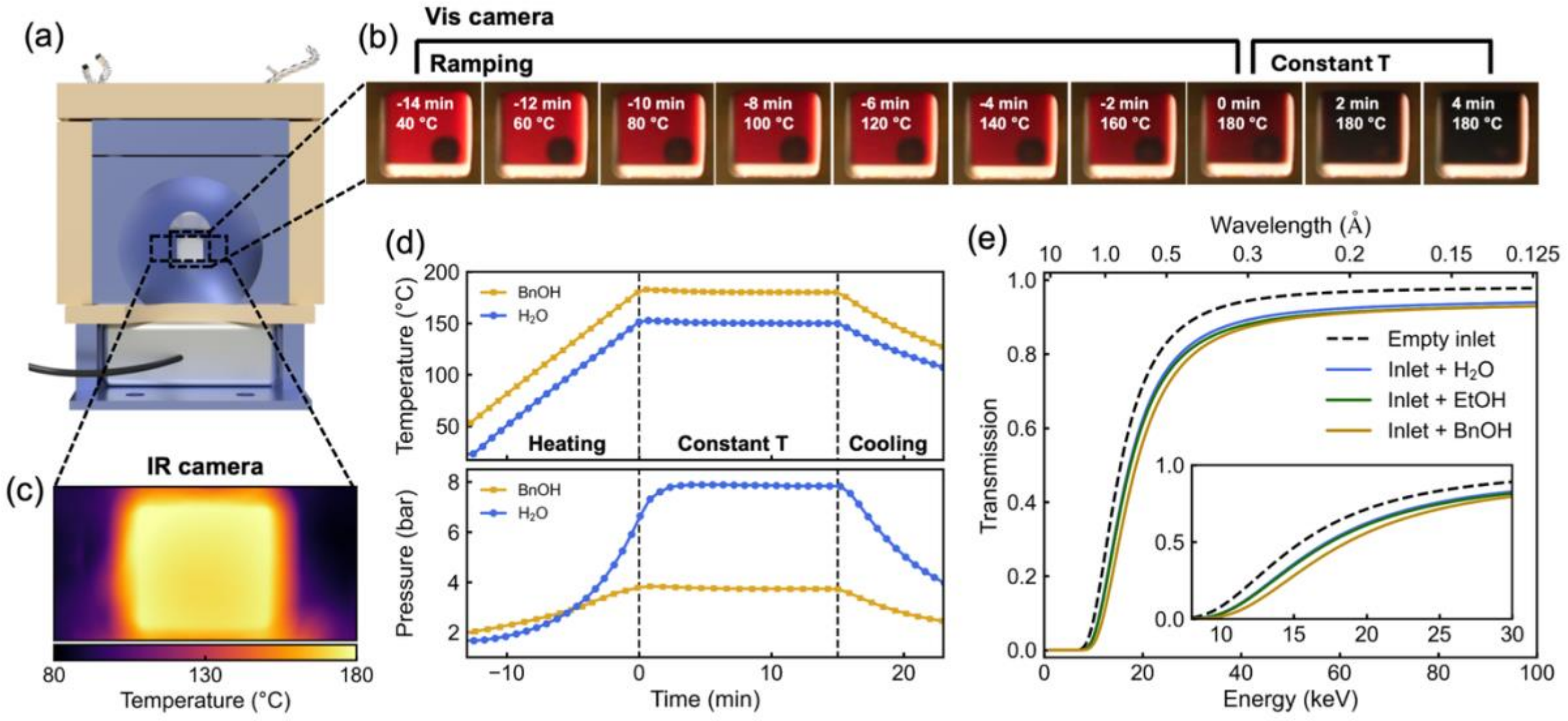

**Figure 2** Thermal, pressure and transmission characteristics of the glass inlet with 0.3 mm wall thickness. (*a*) Front view rendered image of the reactor (*b*) Photographs showing the solution color change during $Fe_3O_4$ formation from $Fe(acac)_3$ in BnOH upon heating. (*c*) IR image showing uniform heat distribution of the BnOH-filled inlet at 180 °C. (*d*) Temperature and pressure profiles during the heating of water and BnOH to 150 °C and 180 °C, respectively. (*e*) Simulated x-ray transmission of the inlet, empty and filled with the solvents water, ethanol (EtOH) and BnOH as a function of the x-ray energy.

The glass inlet configuration offers good pressure stability and airtightness, enabling solvothermal synthesis under autoclave-like conditions for *in situ* scattering experiments. However, the relatively thick inlet walls give rise to a substantial scattering background. One major challenge of *in situ* scattering studies is the strong background signal, as scattering arises from all components in the beam path. To isolate the signal of the chemical species of interest, contributions from the solvent, reactor walls, and beamline environment must be carefully subtracted. In most PXRD experiments, background subtraction is relatively straightforward due to the strong diffraction signal of crystalline phases. However, for weakly scattering systems such as nanoparticles or dilute dispersions or solutions, background subtraction becomes significantly more demanding. This is especially true for total scattering experiments, where often non-periodic structures or small nanoparticles are being investigated and data is collected at high scattering vectors (typically $q_{max} > 15$ Å$^{-1}$), where scattering is weak and noisy. Moreover, standard baseline correction methods are not applicable in TS experiments, as both Bragg and diffuse scattering must be preserved, making accurate and precise background subtraction essential. Figure 3(*a*) illustrates this by comparing PXRD patterns of commercial copper(II) oxide (CuO) nanoparticles (7.4 nm diameter) at concentrations of 0.2, 0.05, and 0.02 mmol/mL in ethanol, highlighting the dominant background relative to the nanoparticle signal. Additionally, we estimate the signal-to-noise ratio (SNR) for each concentration (Figure 3(*b*-*e*)), demonstrating that the reactor with glass inlets can still provide reasonable PXRD data even at very low particle concentrations.

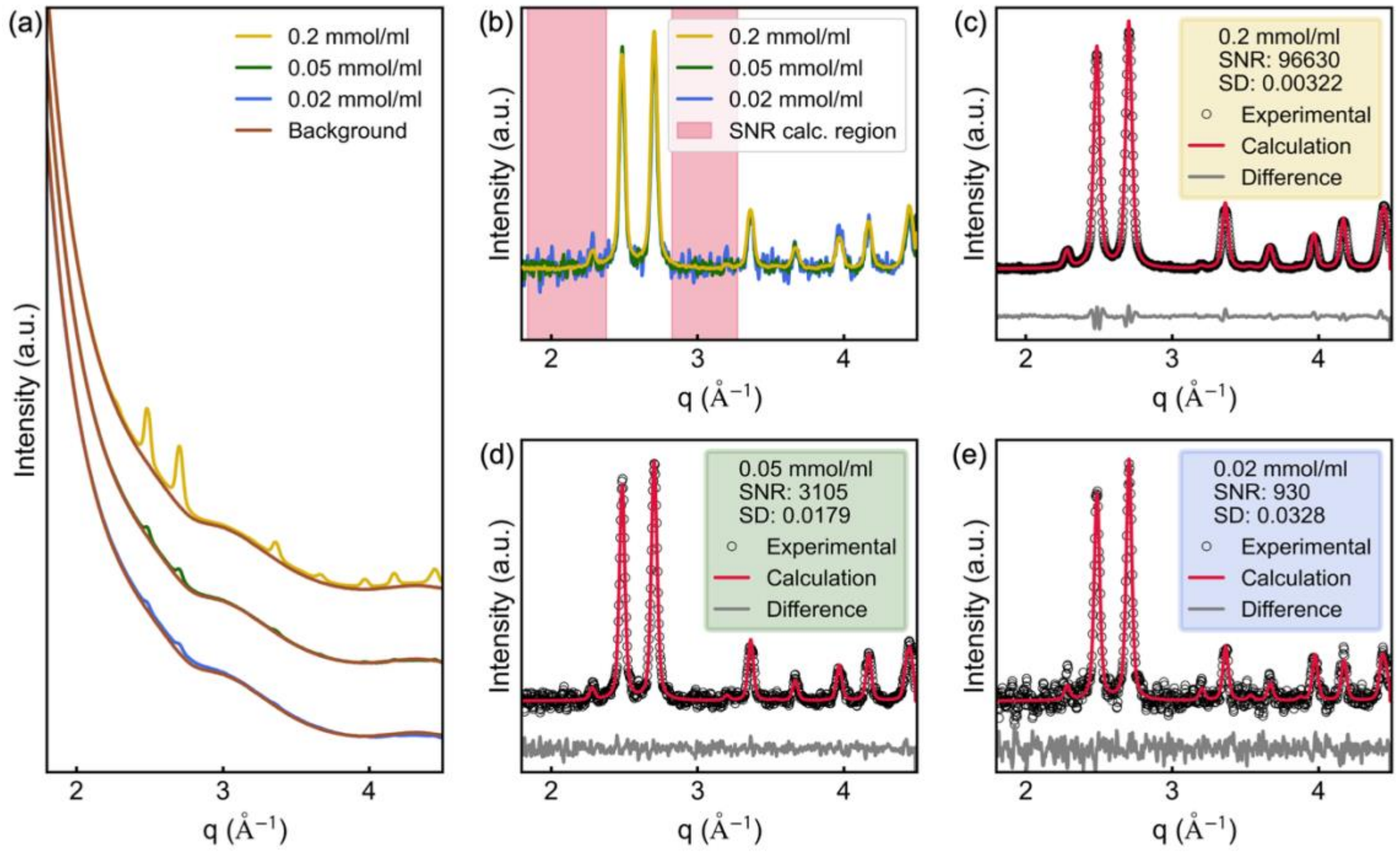

**Figure 3** PXRD of CuO nanoparticles (7.4 nm diameter) dispersed in ethanol at concentrations of 0.2, 0.05, and 0.02 mmol/mL and signal-to-noise ratio (SNR) estimations. (*a*) PXRD of the CuO dispersions compared to the background pattern which is the ethanol filled inlet. (*b*) Overlay of background-subtracted and normalized PXRD patterns for CuO dispersions showing increasing noise with decreasing concentration. (*c-e*) Rietveld refinements for each concentration. To estimate the SNR, we first normalize each pattern to 1 at the most intense Bragg peak (~2.7 $Å^{-1}$). The SNR is then calculated as the inverse square of the standard deviation (SD) of the difference between the experimental data and the Rietveld fit. To avoid influence from imperfections in the fit, regions containing strong Bragg reflections are excluded from the SD calculation as highlighted in (*a*).

We further test the reactor by studying the solvothermal synthesis of $Fe_3O_4$ from $Fe(acac)_3$ in BnOH at 180 °C. (Pinna *et al.*, 2005) This reaction is fast, scalable, and allows tailoring the size, morphology, and even synthesis of heterostructures. (Kim *et al.*, 2009; Fantechi *et al.*, 2017; Nobile and Cozzoli, 2022) The organic reaction pathways during the synthesis of metal-oxides in benzyl alcohol, including $Fe_3O_4$, are well-known and have been assessed in previous studies via gas chromatography-mass spectrometry (GC-MS) and nuclear magnetic resonance (NMR) analysis of the final reaction mixture. In particular, it has been proposed that benzyl acetate and 4-phenyl-2-butanone form via the reaction of benzyl alcohol with $Fe(acac)_3$. Furthermore, during the reaction, one third of Fe(III) species are reduced to Fe(II) by dehydrogenative oxidation of 4-phenyl-2-butanone into 4-phenyl-3-buten-2-one. The complete organic transformation pathway is reported in literature. (Niederberger and Garnweitner, 2006) However, *in situ* studies providing mechanistic insights that focus on the metal center, *e.g.* formation of intermediate complexes and monitoring the nucleation and growth of the nanoparticles, are scarce. (Mikhailova *et al.*, 2022)

Based on our *in situ* scattering data, we propose a reaction mechanism, schematically illustrated in Figure 4(*a*). Initially, $Fe(acac)_3$ is dissolved in BnOH with the Fe(III) center octahedrally coordinated by three acetylacetonate ligands, each coordinating through both oxygen atoms. Upon reaching 180 °C, an intermediate ferric acetate complex, $[Fe_3(\mu_3\text{-}O)(AcOR)_6(ROH)_3]^+$, forms. In this complex, three Fe(III) centers are linked to a central coplanar $\mu_3$-oxo ligand. Acetate groups (AcOR) bridge pairs of Fe(III) centers and are further coordinated by water or alcohol molecules (ROH), maintaining an octahedral coordination. Here, R denotes either a hydrogen atom or a benzyl group. The iron complex with coordinated acetate and water is well known in literature. (Anson *et al.*, 1997) With continued heating, the intermediate transforms into crystalline $Fe_3O_4$ nanoparticles.

Figure 4(*b,c*) show the temperature profiles and heatmaps of the obtained *in situ* PXRD patterns and PDFs, respectively. The reaction solution is ramped at 10 °C/min, with the time point of 0 min defined as the moment when the reaction temperature of 180 °C is reached. Both datasets

clearly reflect the sequential transformation through the three structural stages depicted in Figure 4(*a*). Changes in the scattering signals around 0 min indicate the formation of the intermediate complex, while the appearance of distinct peaks after ~10 min marks the formation of $Fe_3O_4$. Figure 4(*d,e*) show representative PXRD and PDF patterns for the three reaction stages: the initial molecular complex (bottom), the intermediate (middle), and the final $Fe_3O_4$ product (top). Rietveld refinement of the PXRD pattern of the final product confirms the phase purity of $Fe_3O_4$. The non-crystalline nature of the initial and intermediate states excludes conventional PXRD refinement; however, their structures can be analyzed via PDF. The initial state is modeled using the structure of $Fe(acac)_3$ while the intermediate state at 5 min reaction time is best described using a three-phase refinement with $Fe(acac)_3$, $[Fe_3(\mu_3\text{-}O)(AcOR)_6(ROH)_3]^+$, and $Fe_3O_4$ in a 1.00:0.92:0.70 scale ratio. Based on the reported $Fe(acac)_3$ decomposition mechanism in benzyl alcohol forming benzyl acetate (Niederberger and Garnweitner, 2006), we propose the actual intermediate is a ferric acetate complex coordinated with benzyl acetate and benzyl alcohol ligands. However, for the PDF refinement, we employ the literature-reported structure $[Fe_3(\mu_3\text{-}O)(AcO)_6(H_2O)_3]^+$, which contains acetate and water ligands. (Anson *et al.*, 1997) We note that PDF analysis cannot unambiguously distinguish between acetate and benzyl acetate ligands, or between coordinated water and benzyl alcohol, due to the prevalence of short-range interatomic correlations involving Fe and the surrounding BnOH solvent. Nevertheless, the excellent agreement between the modeled and experimental PDFs (Figure 4(*e*)) strongly supports the structural assignment of the intermediate as a ferric acetate complex. The refined parameter of the Rietveld and PDF refinements are given in Table S1 and S2, respectively.

To assess the time resolution achievable with the reactor setup, Figure S5 compares PXRD and PDF data averaged over varying exposure times. These tests demonstrate that high-quality data can be obtained with exposures as short as 5 s during the early, low-scattering stages of the reaction, and down to 1 s for the strongly scattering crystalline final product. Overall, these results highlight the capability of the reactor to provide high-quality PXRD and PDF data at low time resolution despite significant background scattering from the glass components.

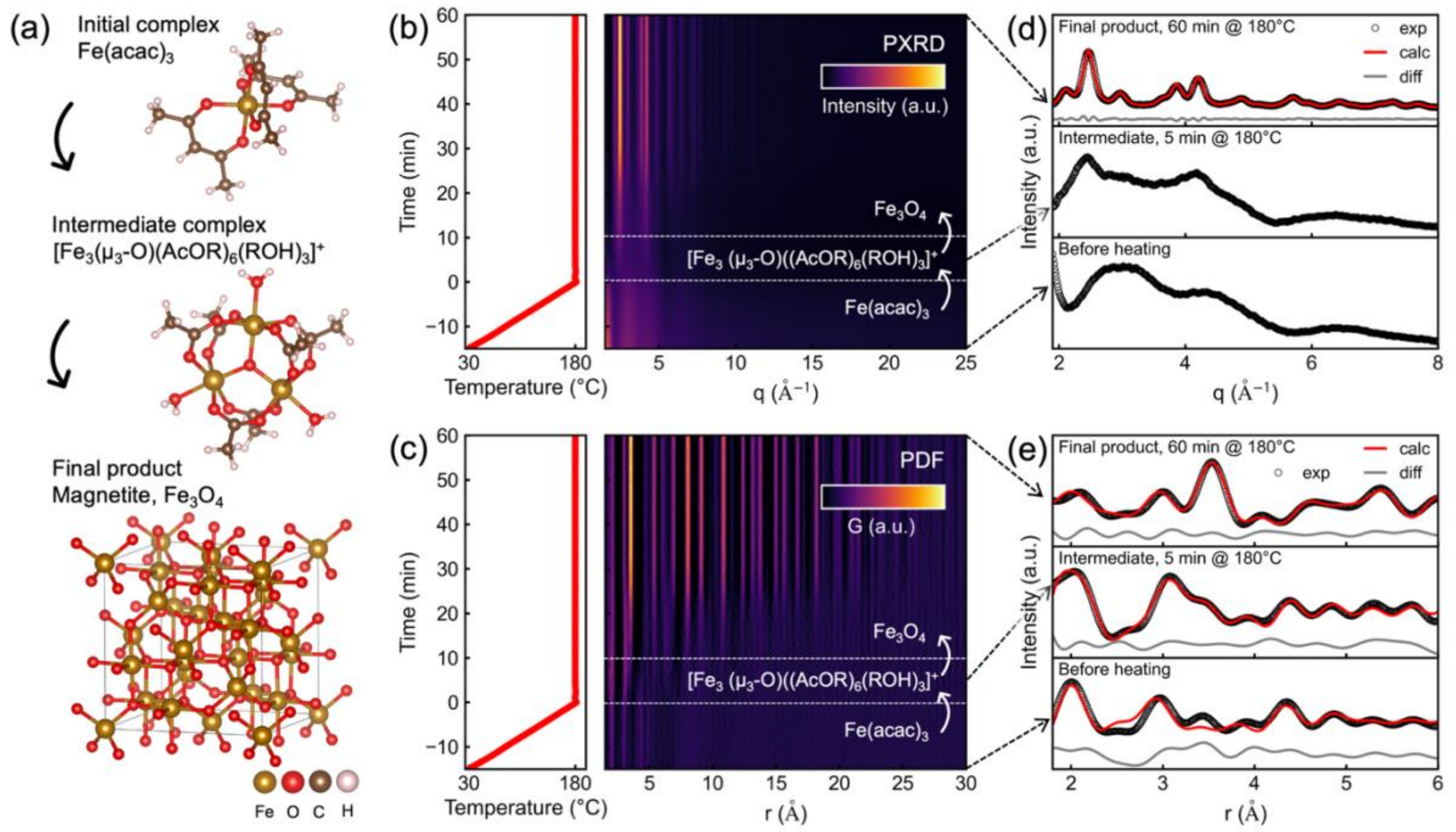

**Figure 4** *In situ* PXRD and PDF analysis of the reaction to $Fe_3O_4$. (*a*) Schematic of the reaction mechanism as determined by PXRD and PDF refinements. The initial complex $Fe(acac)_3$ forms $[Fe_3(\mu_3\text{-}O)(AcOR)_6(ROH)_3]^+$ after reaching the reaction temperature of 180 °C which further reacts to the final product $Fe_3O_4$. (*b,c*) Temperature profile and heatmap of PXRD and PDF data, respectively. (*d,e*) PXRD Rietveld and PDF refinements of the initial complex before heating, $Fe(acac)_3$, the intermediate complex, $[Fe_3(\mu_3\text{-}O)(AcOR)_6(ROH)_3]^+$, at 5 min after reaching the reaction temperature of 180 °C and $Fe_3O_4$. Since Rietveld refinement requires a crystalline structure only the refinement of $Fe_3O_4$ is shown for PXRD.

The PDF refinement of the intermediate state assumes the thermal decomposition of acetylacetonate ligands into acetate. To track the formation of organic by-products in the liquid phase during the reaction, we perform *in situ* ATR Fourier transformed infrared (FTIR) spectroscopy using the reactor equipped with an ATR-IR fiber probe. *In situ* ATR-FTIR difference spectra collected during the reaction and reference spectra for comparison are displayed in Figure 5. Already during the heating process, we observe a decrease in absorption in the frequency region around 1000 $cm^{-1}$, which we assign to the $\nu$(C–O) stretching vibration of BnOH. Furthermore, we observe a decrease in absorption around 1600 $cm^{-1}$, which we attribute to the bending vibration $\delta$(H–O–H) of residual water present in the sample at ambient conditions. As the temperature increases, this water evaporates and transitions into the gas phase, leaving the liquid phase and thus causing the observed signal loss. With further heating, a broadening and increase in absorption around 1225 $cm^{-1}$, which we assign to the emergence of *v*(C–O) vibrations characteristic of an acetate functional group. We correlate this increase with the formation of benzyl acetate, supported by additional vibrational modes around 1739 cm$^{-1}$ (*v*(C=O)), 1026 cm$^{-1}$ and the symmetric and asymmetric bending vibrations (δ($CH_3$)) at 1361 and 1380 cm$^{-1}$, respectively. These observations are consistent with the reported formation of acetates during acetylacetonate decomposition. (Mikhailova *et al.*, 2022) In addition to vibrational modes attributed to benzyl acetate formation, we also identify a vibrational mode at 1717 cm$^{-1}$ (*v*(C=O)), due to the contribution of acetone in the reaction mixture, which is formed alongside benzyl acetate. The assigned vibrational modes are listed in Table S3. For comparison, additional reference spectra are shown in Figure S6. These ATR-FTIR findings for the synthesis of $Fe_3O_4$ support the formation of a ferric acetate complex as an intermediate. They also demonstrate the capabilities of our reactor to acquire ATR-FTIR spectra with a spectral resolution of 4 cm$^{-1}$ and a temporal resolution of less than 1 minute during solvothermal synthesis.

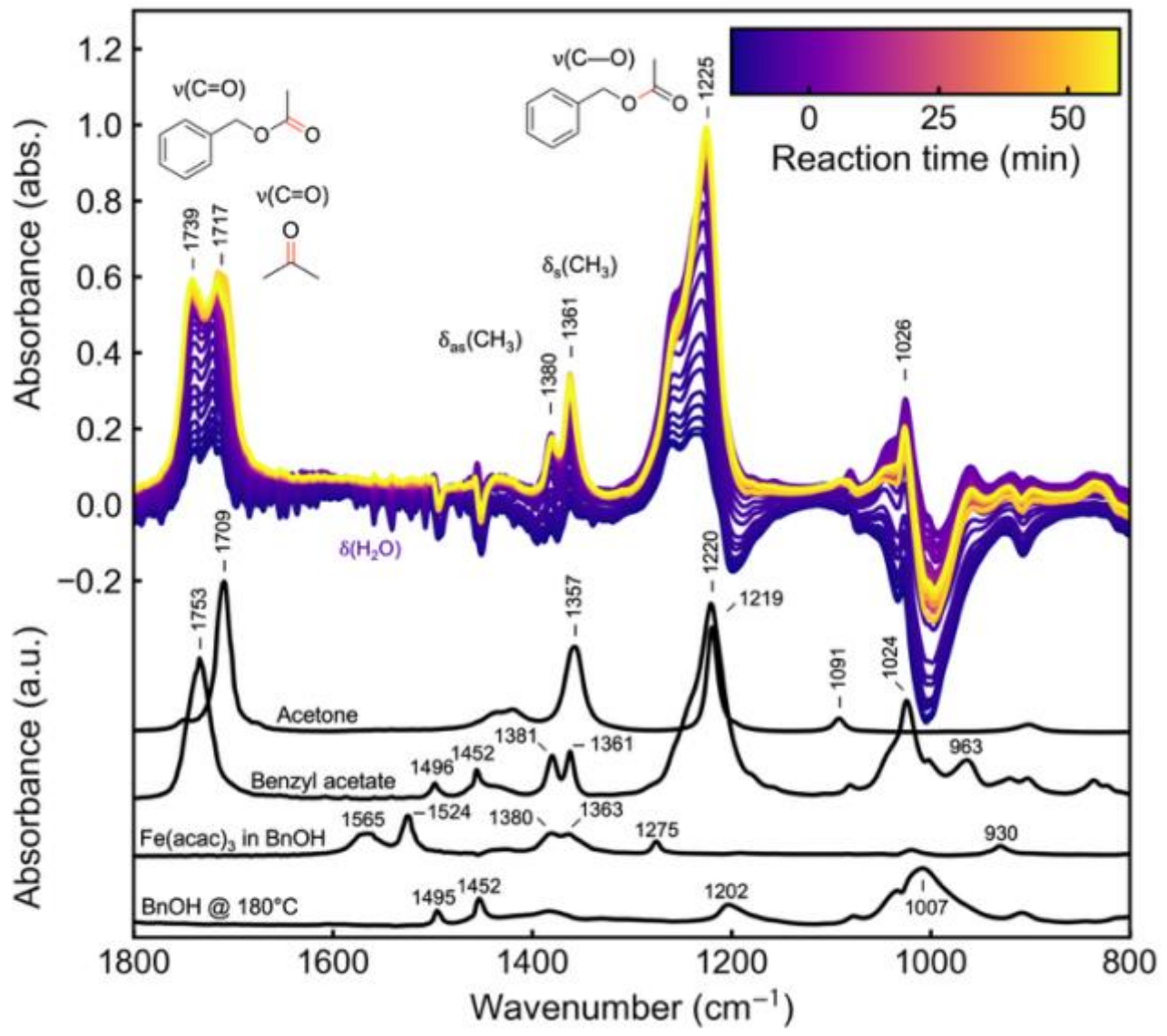

**Figure 5** *In situ* ATR-FTIR analysis of the reaction of $Fe(acac)_3$ to $Fe_3O_4$ in BnOH, compared to reference spectra of acetone and benzyl acetate, $Fe(acac)_3$ dissolved in BnOH at room temperature, and BnOH at 180 °C.

Finally, we investigate the $Fe_3O_4$ solvothermal synthesis using XAS with the PEEK inlet configuration. Figure 6(*a*) presents Fe K-edge high-energy resolution fluorescence-detected XANES (HERFD-XANES) spectra, showing a shift of the absorption edge to lower energy – indicating a partial reduction of $Fe^{3+}$ in the initial and intermediate complexes to a mixed-valence $Fe^{2+}/Fe^{3+}$ state in $Fe_3O_4$. Concurrently, the white-line double peak observed in the initial state transitions into a single, broader feature in the final product, while the pre-edge intensity increases, indicating changes in Fe coordination symmetry and oxidation state. These spectral features closely resemble previously reported XANES spectra of $Fe(acac)_3$ (Bauer *et al.*, 2005; Levish and Winterer, 2020) and $Fe_3O_4$ (Okudera *et al.*, 2012; Piquer *et al.*, 2014), supporting the identification of the initial complex as $Fe(acac)_3$ and confirming the formation of $Fe_3O_4$ as the final product.

To analyze these changes quantitatively, we perform multivariate curve resolution by alternating least squares (MCR-ALS) (Jaumot *et al.*, 2005) on the *in situ* HERFD-XANES dataset, which extracts three distinct Fe species, consistent with the PDF findings. Figure 6(*b*) shows the evolution of the three species during the reaction. Figure 6(*c*) shows the MCR-ALS recovered spectra. In Figure 6(*d*) we present results from FEFF simulations (Rehr *et al.*, 2010) based on the structures determined from the PDF analysis. Since $Fe_3O_4$ comprises octahedral and tetrahedral sites in a 2:1 ratio, the simulation of $Fe_3O_4$ shows the linear combination of FEFF simulations with the octahedral and tetrahedral Fe as the absorbing atom in a 2:1 ratio. Since $Fe_3O_4$ contains Fe atoms in both octahedral and tetrahedral coordination sites in a 2:1 ratio, the simulation of $Fe_3O_4$ was performed as a linear combination of FEFF simulated spectra calculated with Fe in octahedral and tetrahedral sites as the absorbing atom, weighted accordingly at a 2:1 ratio. The simulations match well with the MCR-ALS recovered spectra. Figure S7 shows the FEFF simulated XANES spectra together with density of state (DOS) calculation of the Fe, O, and C s, p ,and d states.

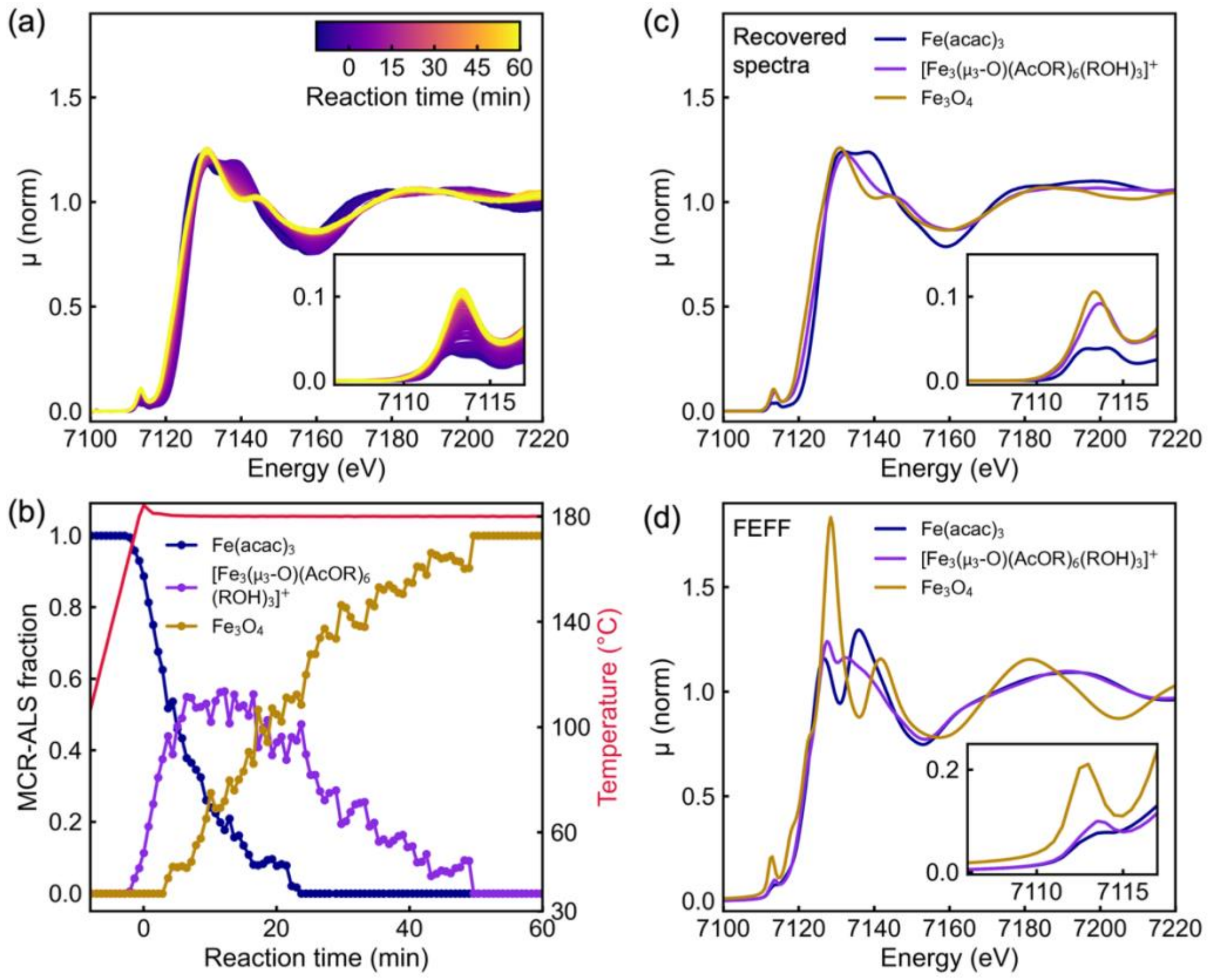

**Figure 6** *In situ* Fe K-edge HERFD-XANES analysis of the reaction to $Fe_3O_4$. (*a*) *In situ* HERFD-XANES data. (*b*) Concentration profiles of three distinct Fe species extracted via MCR-ALS. (*c*) MCR-ALS recovered spectra. (*d*) FEFF simulations based on structures determined from in situ PDF analysis, confirming the initial species as $Fe(acac)_3$, the intermediate as $[Fe_3(\mu_3\text{-}O)(AcOR)_6(ROH)_3]^+$, and $Fe_3O_4$ as the final product.

Overall, these results showcase the excellent performance of the reactor for *in situ* analyses such as high-resolution XAS and WAXS, as well as ATR-FTIR measurements, and demonstrate its capability to provide detailed electronic and structural insight during solvothermal synthesis.

## 4. Conclusions

We present a versatile reactor system optimized for *in situ* x-ray scattering and x-ray absorption spectroscopy analysis of solvothermal reactions. The reactor supports techniques such as XAS, WAXS, and SAXS under inert, autoclave-like conditions. Its modular design features exchangeable inlets tailored to specific experimental requirements: PEEK for x-ray spectroscopy, and glass inlets for x-ray scattering. Additional configurations enable gas or liquid injection and integration of an ATR-IR fiber probe for simultaneous ATR-FTIR measurements. The reactor enables precise temperature control (-20 °C to 200 °C) with cartridge heaters or Peltier elements and magnetic stirring. The compact geometry is compatible with synchrotron beamline constraints and supports data acquisition in both transmission and reflection geometries.

We demonstrate the reactor's performance by investigating the synthesis of $Fe_3O_4$ nanoparticles from $Fe(acac)_3$ in benzyl alcohol as shown in Figure 7. By linking findings from *in situ* PXRD, PDF of TS, HERFD-XANES, and ATR-FTIR, we comprehensively study the reaction pathways and reveal ferric acetate as an intermediate structure. This study underlines the potential of our reactor platform to provide mechanistic insights into solvothermal reactions and nanoparticle formation processes in real time, paving the way for rational synthesis design.

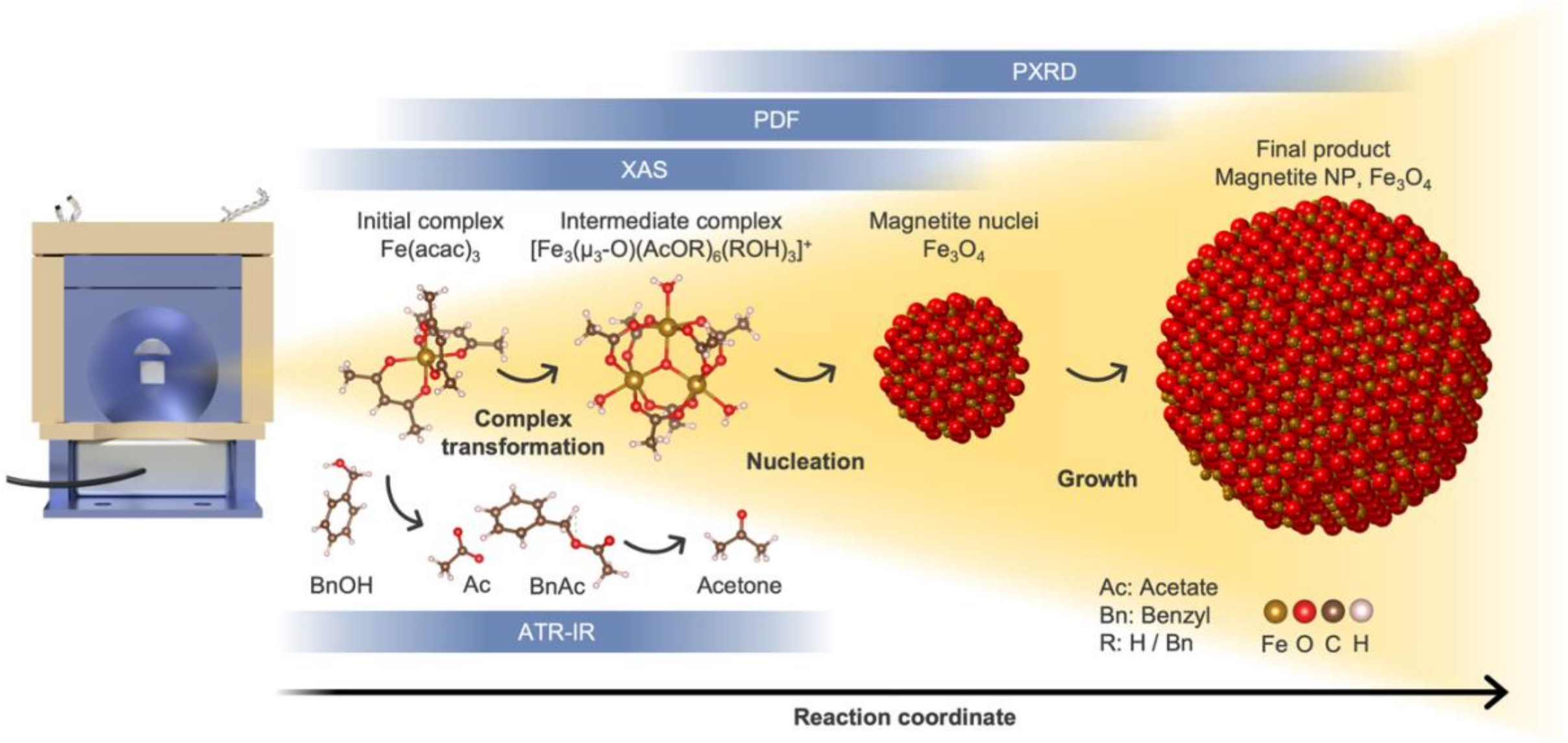

**Figure 7** Schematic overview of the reaction pathway leading to magnetite ($Fe_3O_4$) nanoparticles (NPs). The mechanism is elucidated using a combination of *in situ* techniques made possible by the versatile reactor setup. ATR-IR spectroscopy monitors the evolution of organic species throughout the reaction. XAS and PDF analyses reveal the conversion of the initial iron(III) acetylacetonate complex to the intermediate ferric acetate complex, followed by the formation of the magnetite phase. PXRD measurements track the emergence of the magnetite crystal structure and nanoparticle growth.

## 5. Experimental section

*Chemicals*

All Chemicals were purchased from commercial sources and used without further purification: $Fe(acac)_3$ (Sigma Aldrich, 99.9%), benzyl alcohol (Sigma Aldrich, 99.8%), copper(II) oxide (Sigma Aldrich, 99.999%), ethanol (VWR chemicals, >99.9%). $Fe(acac)_3$ and benzyl alcohol were stored and handled in the glove box under inert atmosphere (Ar 6.0 purity, $c(H_2O) < 0.1$ ppm, $c(O_2) < 0.1$ ppm).

*Synthesis*

In a typical synthesis, a stock solution is prepared in the glove box by adding $Fe(acac)_3$ (353.2 mg, 1.0 mmol) to 5 mL of benzyl alcohol and stirring until the iron salt was completely dissolved. Subsequently, a quantity of 80 µL was transferred to the inlet of the reactor. After assembling the reactor, it was taken out of the glovebox and heated to 180 °C with a heating rate of 10 °C/min under vigorous stirring. After the desired reaction time at 180 °C the reactor was cooled to room temperature.

*X-ray transmission calculations*

X-ray transmission was calculated by first determining the X-ray absorption coefficients of the respective materials using the *Xraydb* Python library (Elam, Ravel and Sieber, 2002; Newville, 2025) and applying Lambert's law to compute the transmission.

*Attenuated Total Reflection Fourier Transformed Infrared (ATR-FTIR) Spectroscopy*

A ThermoFisher Nicolet iS20 equipped with a liquid $N_2$-cooled MCT detector, combined with a 6.3 mm high-temperature diamond ATR-probe (art photonics, Berlin, Germany), was used for *in situ* ATR-IR measurements. Difference spectra were acquired using 32 scans and a spectral resolution of 4 $cm^{-1}$, resulting in a time resolution of 48 s. To minimize thermal effects on the absorbance, a background of benzyl alcohol preheated to 180 °C was used for the *in situ* analytical experiments. Spectra of reference compounds were acquired in absolute mode with a background of fiber in air.

*In situ High-Energy Resolution Fluorescence Detected X-ray Absorption Near Edge Structure (HERFD-XANES)*

The *in situ* HERFD-XAS data was acquired at the ID26 beamline (Gauthier *et al.*, 1999) at the European Synchrotron Radiation Facility (ESRF) in Grenoble, France. HERFD-XAS measurements were performed by measuring the intensity of the Fe Kα main line using a Ge(440) crystal in Rowland geometry (Glatzel *et al.*, 2021) while scanning the incident energy across the range of 7.10 to 7.22 keV with a step size of 0.1 eV. An Si(111) monochromator was used, and the overall energy resolution was ~1.4 eV. To minimize radiation damage, the incident X-ray beam spot is moved onto the reaction cell after each scan, as shown in Figure S9 and S10. Each HERFD-XAS spectrum was collected over 35 seconds.

The HERFD-XAS datasets were processed utilizing a custom Python script. The absorption edge position was determined, and the edge jump was normalized using the LARCH-XAFS software module. (Newville, 2013) Spectroscopic data underwent smoothing via a Savitzky-Golay filter and additional processing with the NumPy and SciPy libraries. A comparisson between the raw and processed data is depicted in Figure S11.

*In situ Pair Distribution Function (PDF) analysis of In Situ Total Scattering (TS) and Powder X-ray Diffraction (PXRD)*

*In situ* TS and PXRD data were acquired at beamline P21.1 at PETRA III at Deutsches Elektronen-Synchrotron DESY, Hamburg, Germany. (v. Zimmermann *et al.*, 2025) Two-dimensional scattering patterns were recorded every 1 s at an x-ray energy of 101.39 keV ($\lambda = 0.1222$ Å) using an x-ray area detector (PerkinElmer XRD1621, Varex Imaging Corp.) with 2048x2048 pixels and a pixel size of $200 \times 200$ μm². A $LaB_6$ powder standard packed into the fused silica inlet of the *in situ* reactor was used to calibrate the sample-to-detector distance: 0.390 m for *in situ* TS measurements of $Fe_3O_4$ synthesis and 1.541 m for ex situ PXRD of CuO. For the TS data, $q_{damp}$ and $q_{broad}$ values were calibrated as 0.0494 Å$^{-1}$ and 0.0374 Å$^{-1}$, respectively. PDF data were processed with $q_{min} = 1.1$ Å$^{-1}$, $q_{max} = 14.2$ Å$^{-1}$, $q_{max,inst} = 24.0$ Å$^{-1}$, and $r_{poly} = 0.9$. Azimuthal integration and calibration were carried out using pyFAI (Kieffer and Wright, 2013), and PDF refinements were performed using diffpy-CMI (Juhás *et al.*, 2015). Rietveld refinements were conducted with GSAS-II (Toby and Von Dreele, 2013). A detailed description of the data processing and PDF analysis procedures is available in the literature. (Harouna-Mayer *et al.*, 2025)

**Acknowledgements** S. Y. Harouna-Mayer and M. G. Akcaalan contributed equally to this work. This research was supported by the European Research Council (LINCHPIN project, grant no. 818941), the Deutsche Forschungsgemeinschaft (DFG) through the Cluster of Excellence "Advanced Imaging of Matter" (EXC 2056, project ID 390715994) and by the Bundesministerium für Bildung und Forschung (BMBF) via the project 05K22GU7 (LUCENT II), and the project 05K2020-2019-06104 XStereoVision (grant no. 05K20GUA). The authors acknowledge DESY (Hamburg, Germany), a member of the Helmholtz Association, and ESRF (Grenoble, France) for the provision of experimental facilities. Parts of this research were carried out at PETRA III using beamline P21.1 (v. Zimmermann *et al.*, 2025) and at ESRF using beamline ID26 under proposal MA5366 (Caddeo *et al.*, 2024). The authors also thank the beamline staff for the support with the experiments: at ID26, Dr. Pieter Glatzel; at P21.1, Dr. Martin v. Zimmermann, Dr. Fernando Igoa, Dr. Jiatu Liu, Philipp Glaevecke, and Olof Gutowski. Further, the authors thank Stefan Fleig and the workshop team at the University of Hamburg for machining the custom reactor components.

**Conflicts of interest** The authors declare that there are no conflicts of interest.

**Data availability** All data presented in this report and three-dimensional step-files of the *in situ* reactor are available at: DOI: 10.25592/uhhfdm.17624. A self-written Python script for operating the LS335 temperature controller is available at https://gitlab.rrz.uni-hamburg.de/koziej-lab/in-situ-cell.

# Supporting information

The reactor was adapted to the cooling application and used with an injection cap for reactions operating at sub-zero temperatures and under inert conditions. For the updated version, the essential modification was to incorporate the Peltier element (Adaptive-ETC-128-14-06-E) into the design to cool the system temperature down to -20 °C. The cooling places further requirements, like cooling the hot side of the Peltier element to avoid overheating and achieve better performance. This requirement is met by placing water-cooled copper blocks next to the Peltier elements by taking advantage of copper's excellent thermal conductivity. The water cooling that cycles inside the copper blocks was provided with a chiller (Hei-Chill series-Heidolph) set at 5 °C. As in the previous design, the reactor temperature was monitored using a sensor mounted on the brass body, whereas the temperature of the Peltier element was controlled by a thermoelectric cooler (TEC) (Meerstetter Engineering TEC-1167) through a side-mounted sensor embedded in the copper block.

In terms of thermal insulation, polypropylene (PP) insulators are positioned on the front, back, and top sides of the brass housing to decrease the thermal leakage from the system to the air.

As for the inlet preference, both glass and PEEK inlets can be used in the reactor. An $N_2$ connection, which is embedded in the housing, blows vertically through the inlet's X-ray window. Thereby, possible frost that can negatively affect the measurement is prevented by displacing the air moisture from the window surface. Figure S1 presents the configuration where the Peltier cooling-incorporated reactor is used with the glass inlet equipped with a brass injection cap.

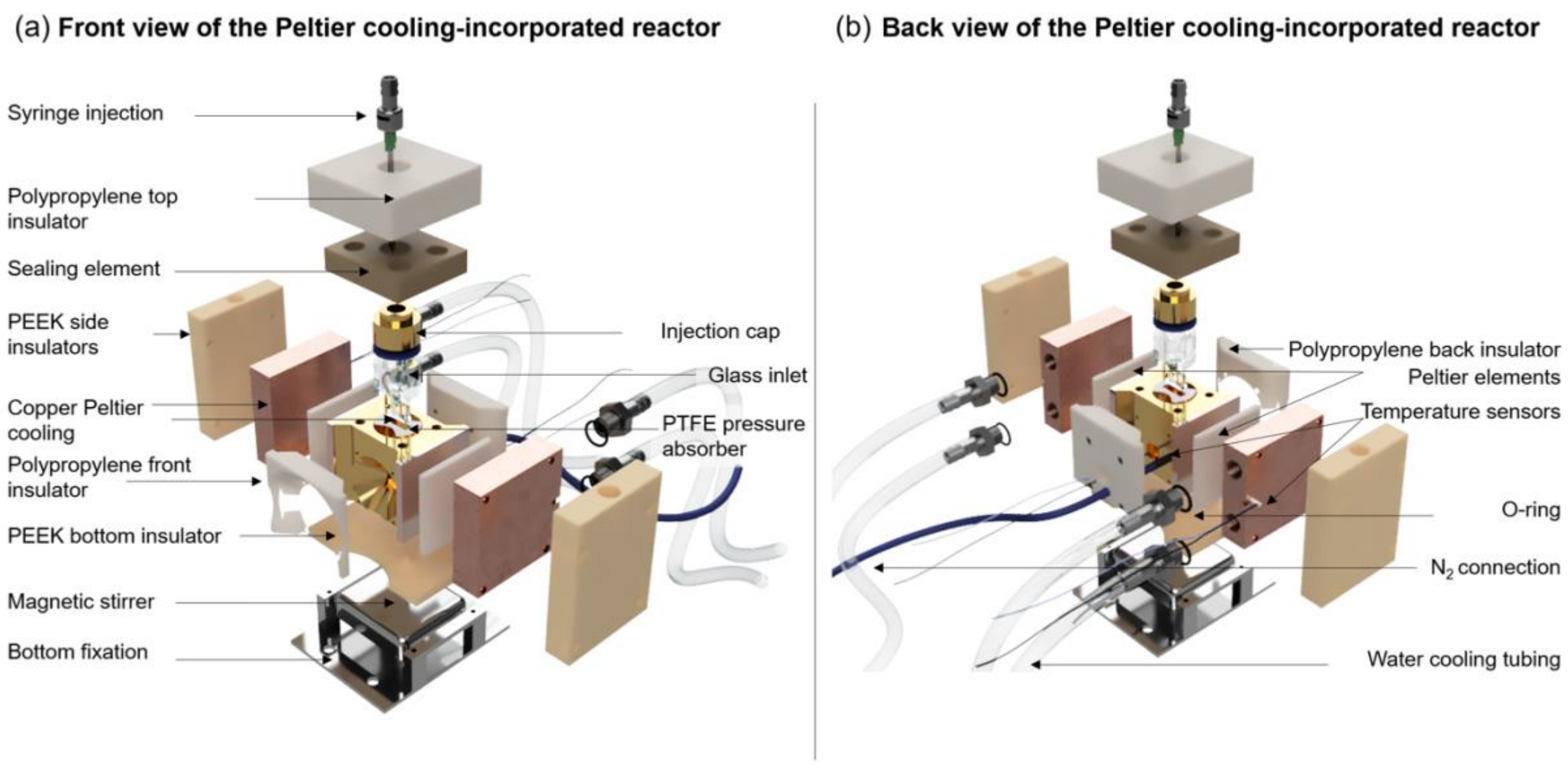

**Figure S1** The rendered image of the Peltier cooling-incorporated reactor. *(a)* Exploded front view. *(b)* Exploded back view.

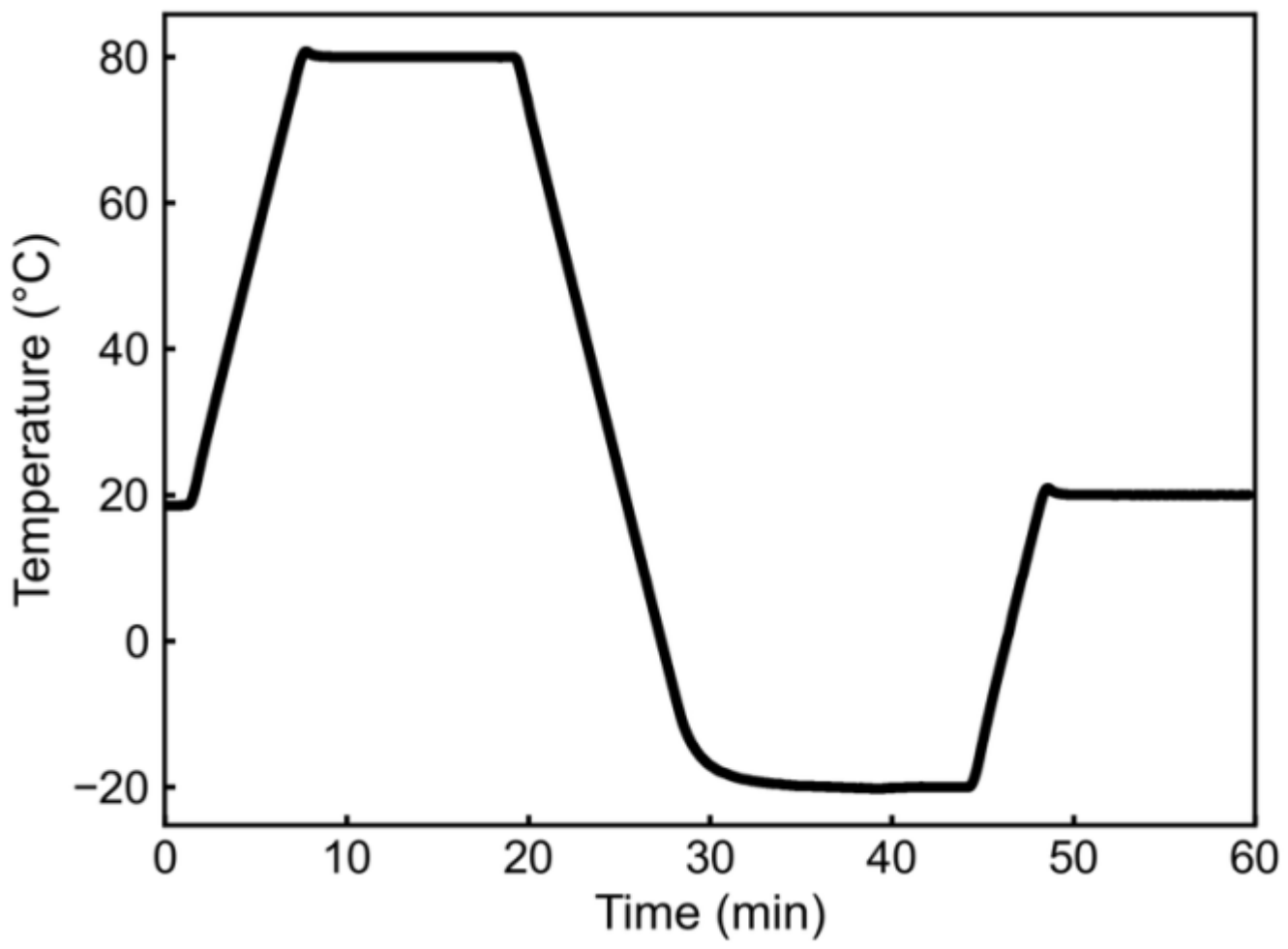

**Figure S2** Temperature profile of the reactor equipped with Peltier cooling, showing the heating ramp to 80 °C followed by active cooling to -20 °C.

As mentioned earlier, the IR-adapted cap can be used with an ATR-IR fiber cable, allowing the cable tip to contact the solution directly and get adequate data. The IR cap consists of two components, a hollow brass screw fastener and a hollow PEEK component. The fiber cable passes through the hollow brass fastener and is screwed to the hollow PEEK cap, and these pieces are sealed into the inlet via an O-ring. Similarly, the pressure sensor is also threaded into a hollow PEEK cap, allowing the sensor to measure the pressure close to the solution interface.

The pressure sensor head is wrapped in PTFE tape first to avoid any loose connection that can affect the pressure measurement. Then, it is screwed to the cap and sealed through an O-ring to the inlet.

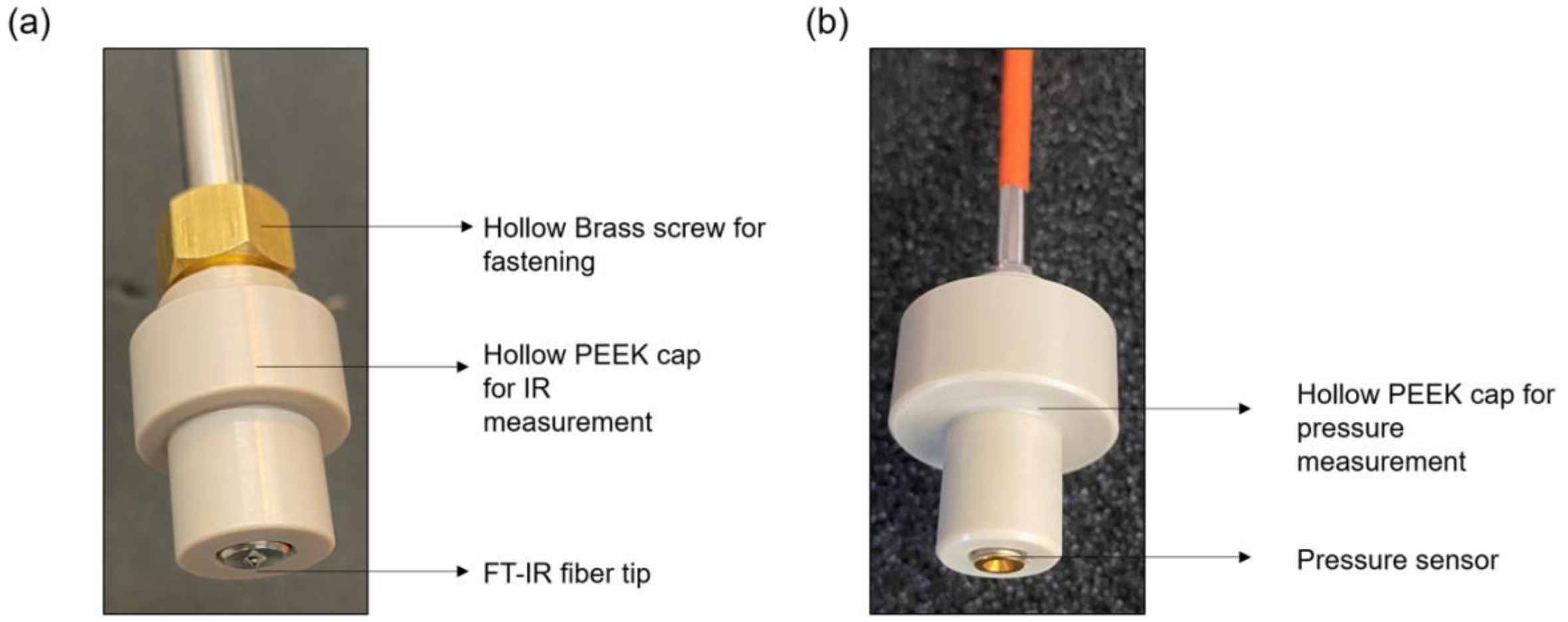

**Figure S3** *(a)* The picture of Brass + PEEK cap equipped with an optical ATR-FTIR fiber probe. *(b)* The picture of a PEEK cap with a pressure sensor.

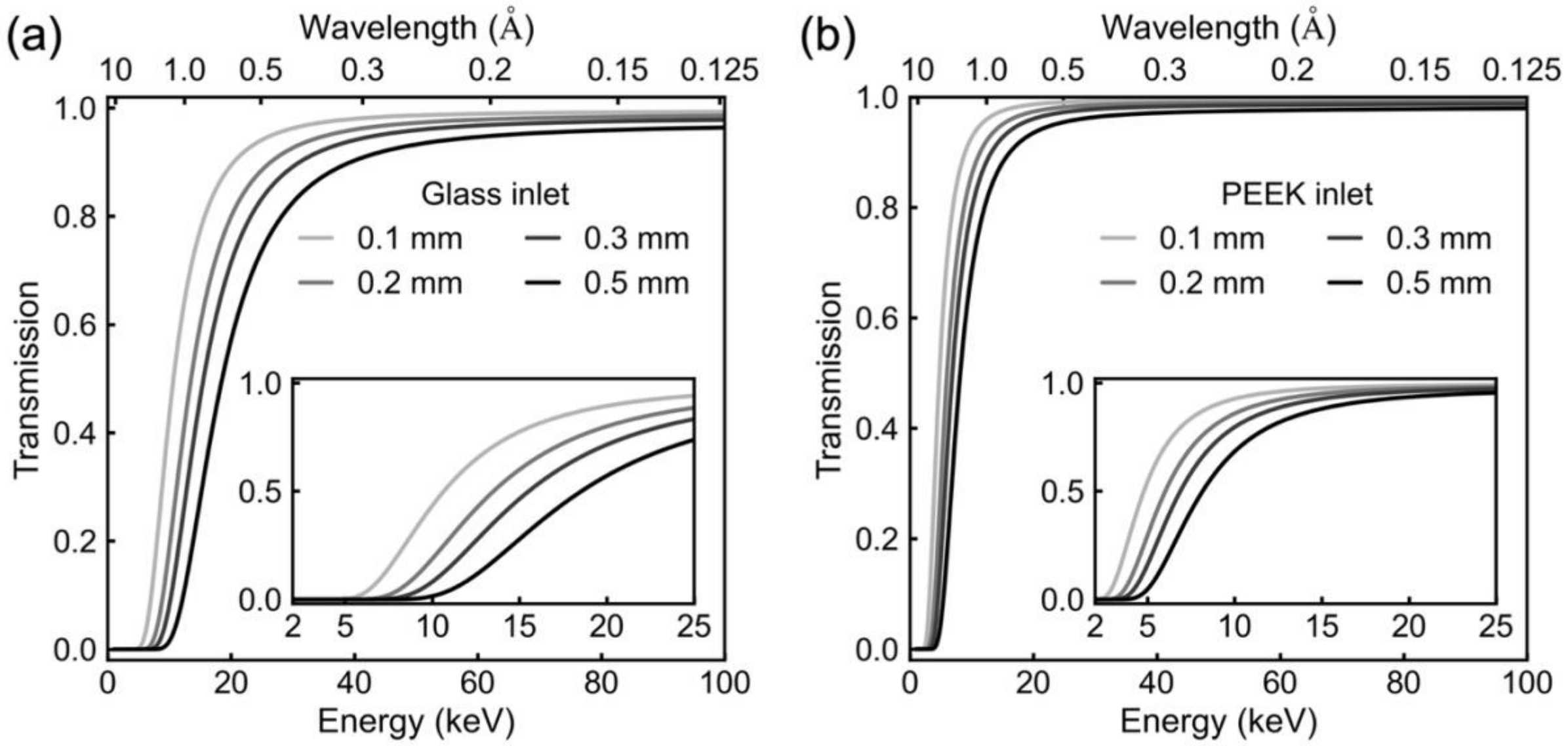

**Figure S4** X-ray transmission of (*a*) the glass and (*b*) PEEK inlet at wall thicknesses between 0.1 and 0.5 mm. The inlet shows a zoom between 2 and 25 keV x-ray energy.

**Table S1** Refined structural parameters from Rietveld analysis of the final stage of the reaction.

| | |
|---|---|
| $R_{wp}$ (%) | 4.34 |
| a (Å) | 8.449 |
| Size (nm) | 4.3 |
| $Fe^{3+}$ $U_{iso}$ (Å$^2$) | 0.0130 |
| $Fe^{2+}$ $U_{iso}$ (Å$^2$) | 0.0096 |
| $O^{2-}$ $U_{iso}$ (Å$^2$) | 0.0066 |

**Table S2** Refined structural parameters from PDF analysis for the initial, intermediate, and final stage of the reaction.

| | Reaction stage | Initial stage | Intermediate stage | Final stage |
|---|---|---|---|---|
| | Reaction time | -15 min | 5 min | 75 min |
| | $R_W$ (%) | 43.9 | 19.0 | 19.8 |
| $Fe(acac)_3$ | Relative scale | 1.000 | 1.000 | 0.000 |
| | $\delta_2$ (Å$^2$) | 1.591 | 1.591 | - |
| | Fe $U_{iso}$ (Å$^2$) | $9.530 \times 10^{-4}$ | $2.077 \times 10^{-4}$ | - |
| | O $U_{iso}$ (Å$^2$) | $1.506 \times 10^{-2}$ | $1.000 \times 10^{-1}$ | - |
| | $C_1$ $U_{iso}$ (Å$^2$) | $3.354 \times 10^{-2}$ | $9.211 \times 10^{-3}$ | - |
| | $C_2$ $U_{iso}$ (Å$^2$) | $1.000 \times 10^{-1}$ | $2.165 \times 10^{-2}$ | - |
| | $C_3$ $U_{iso}$ (Å$^2$) | $1.966 \times 10^{-3}$ | $9.961 \times 10^{-2}$ | - |
| $[Fe_3O(AcO)_6(H_2O)_3]^+$ | Relative scale | - | 0.922 | - |
| | $\delta_2$ (Å$^2$) | - | 3.369 | - |
| | Fe $U_{iso}$ (Å$^2$) | - | $1.365 \times 10^{-2}$ | - |
| | O $U_{iso}$ (Å$^2$) | - | $1.000 \times 10^{-1}$ | - |
| | C $U_{iso}$ (Å$^2$) | - | $3.399 \times 10^{-2}$ | - |
| $Fe_3O_4$ | Relative scale | 0.000 | 0.703 | 1.000 |
| | a (Å) | - | 8.859 | 8.437 |
| | $\delta_2$ (Å$^2$) | - | 2.370 | 0.823 |
| | Fe $U_{11}$ (Å$^2$) | - | $1.783 \times 10^{-1}$ | $8.782 \times 10^{-3}$ |
| | Fe $U_{12}$ (Å$^2$) | - | $5.192 \times 10^{-4}$ | $1.645 \times 10^{-3}$ |

| | | | |
|---|---|---|---|
| Fe $U_{iso}$ ($Å^2$) | - | 6.333 x$10^{-3}$ | 2.466 x$10^{-3}$ |
| O $U_{11}$ ($Å^2$) | - | 2.610 x$10^{-1}$ | 1.622 x$10^{-2}$ |
| O $U_{12}$ ($Å^2$) | - | 2.626 x$10^{-2}$ | 1.413 x$10^{-2}$ |

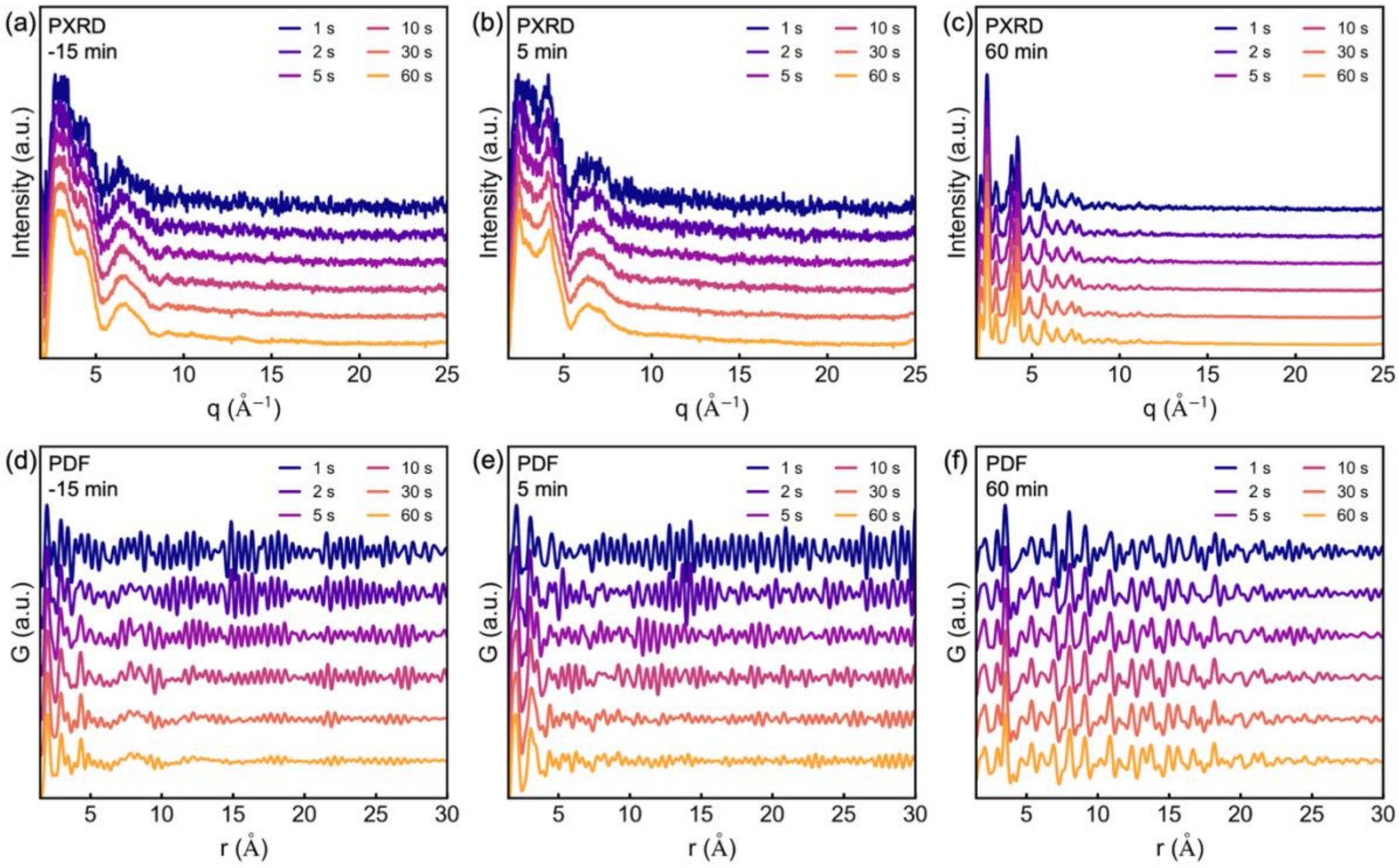

**Figure S5** PXRD and PDF data collected at –15, 5, and 60 minutes reaction time. Data are averaged over 1 to 60 individual patterns, each acquired with a 1 s exposure time.

**Table S3** Assigned vibrational modes observed in the *in situ* ATR-FTIR spectra.

| Peak position ($cm^{-1}$) | Vibrational mode | Assigned to |
|---|---|---|
| 1739 | ν(C=O) | Benzyl acetate |
| 1717 | ν(C=O) | Acetone |
| 1380 | $\delta_{as}(CH_3)$ | Methyl |
| 1361 | $\delta_s(CH_3)$ | Methyl |
| 1225 | ν(C–O) | Benzyl acetate |

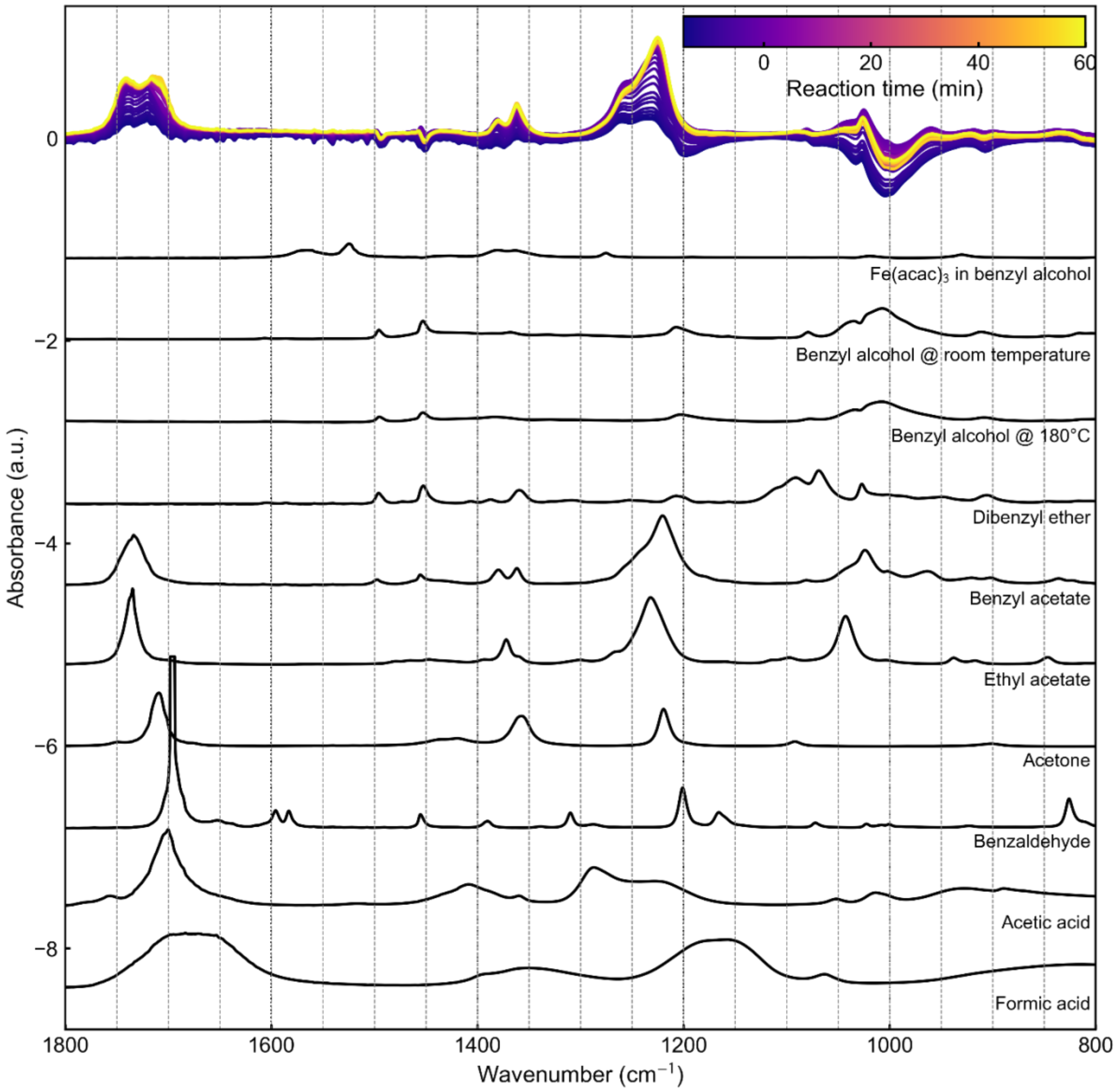

**Figure S6** *In situ* ATR-FTIR analysis of the reaction of $Fe(acac)_3$ to $Fe_3O_4$ in BnOH, compared to reference spectra of $Fe(acac)_3$ dissolved in BnOH, BnOH at room temperature, and at 180 °C, dibenzyl ether, benzyl acetate, ethyl acetate, acetone, benzaldehyde, acetic acid, and formic acid.

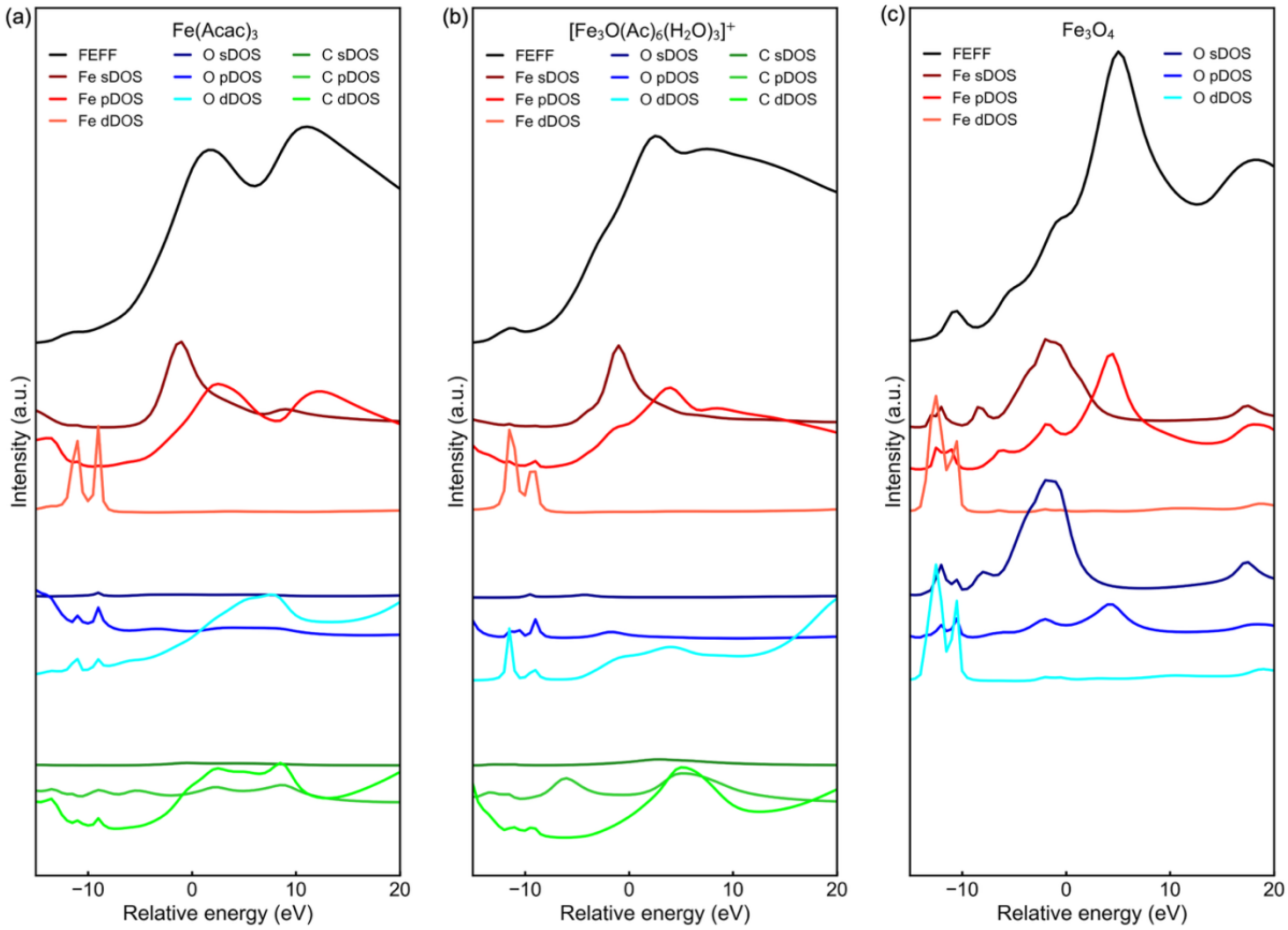

**Figure S7** FEFF simulated XANES spectra and their corresponding density of states (DOS) of s, p and d states of (*a*) $Fe(acac)_3$, (*b*) $[Fe_3O(AcO)_6(H_2O)_3]^+$, and (*c*) $Fe_3O_4$.

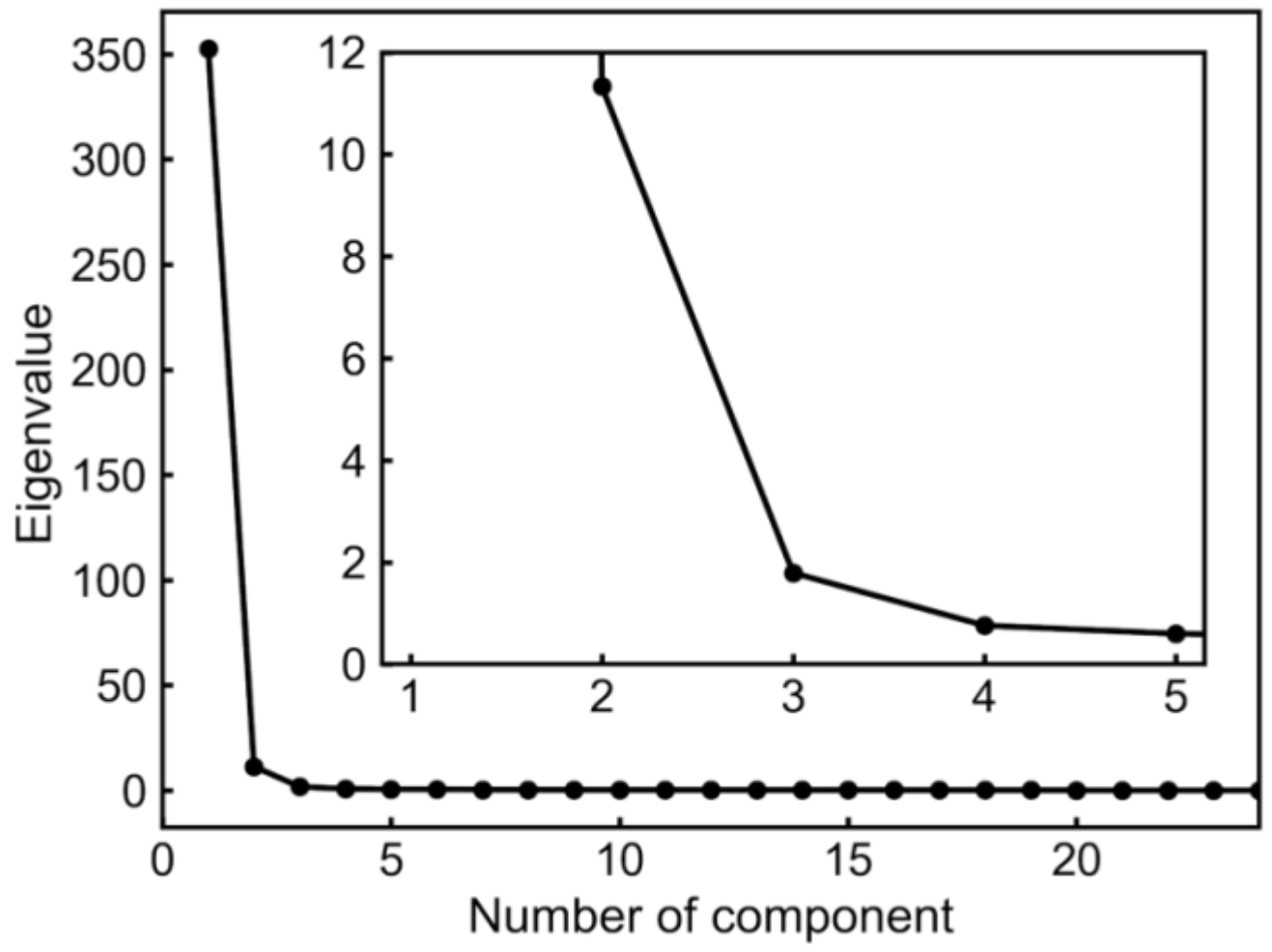

**Figure S8** Eigenvalue profile of MCR-ALS analysis.

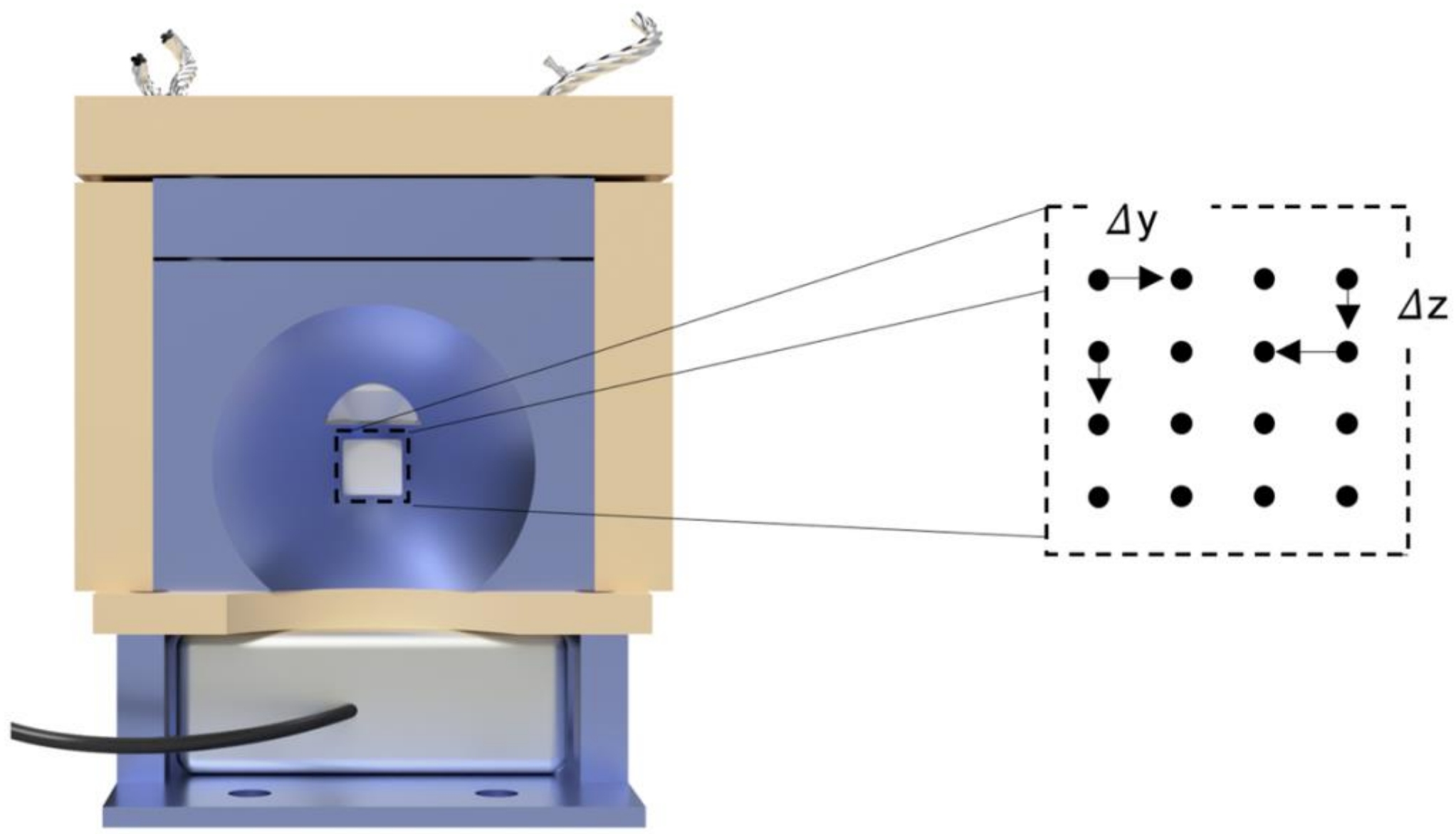

**Figure S9** Schematic of the scanning pattern across the x-ray window used for XAS measurements to minimize beam damage. Repeated exposure of the same sample area is avoided by collecting spectra at different sample positions along a defined trajectory with step sizes Δy and Δz along the horizontal and vertical axes of the sample window.

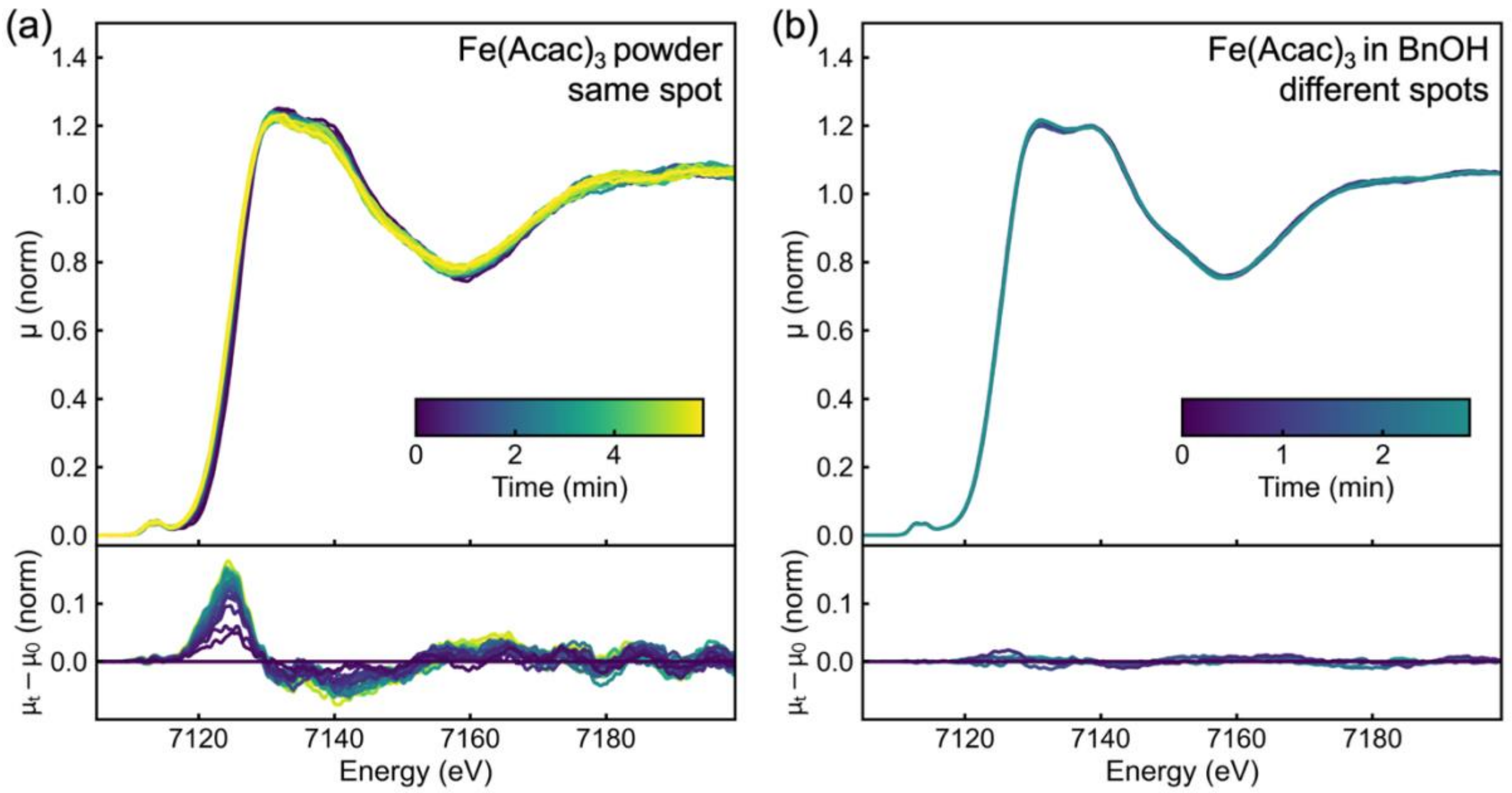

**Figure S10** Beam damage study on $Fe(acac)_3$. (*a*) Fe K-edge HERFD-XANES spectra of $Fe(acac)_3$ powder recorded at room temperature over 30 consecutive scans (12 s each) at the same sample position on the sample. (*b*) Fe K-edge HERFD-XANES data of $Fe(acac)_3$ in benzyl alcohol (BnOH) recorded at room temperature over 5 consecutive scans (42s each) at different sample positions. The upper panels in (*a*) and (*b*) show the HERFD-XANES spectra while the lower panels show the difference of the spectra at the respective time $\mu_t$ and the first spectra at 0 min $\mu_0$. The spectra recorded at the same sample spot (*a*) show beam damage evident by the shift of the absorption edge and the decrease of the white-line intensity. The spectra recorded at different sample spots (*b*) show no strong spectral changes. This demonstrates that scanning across different sample positions, as illustrated in Figure S9, effectively enables avoiding beam damage. These scanning conditions were used for the *in situ* Fe K-edge HERFD-XANES measurements.

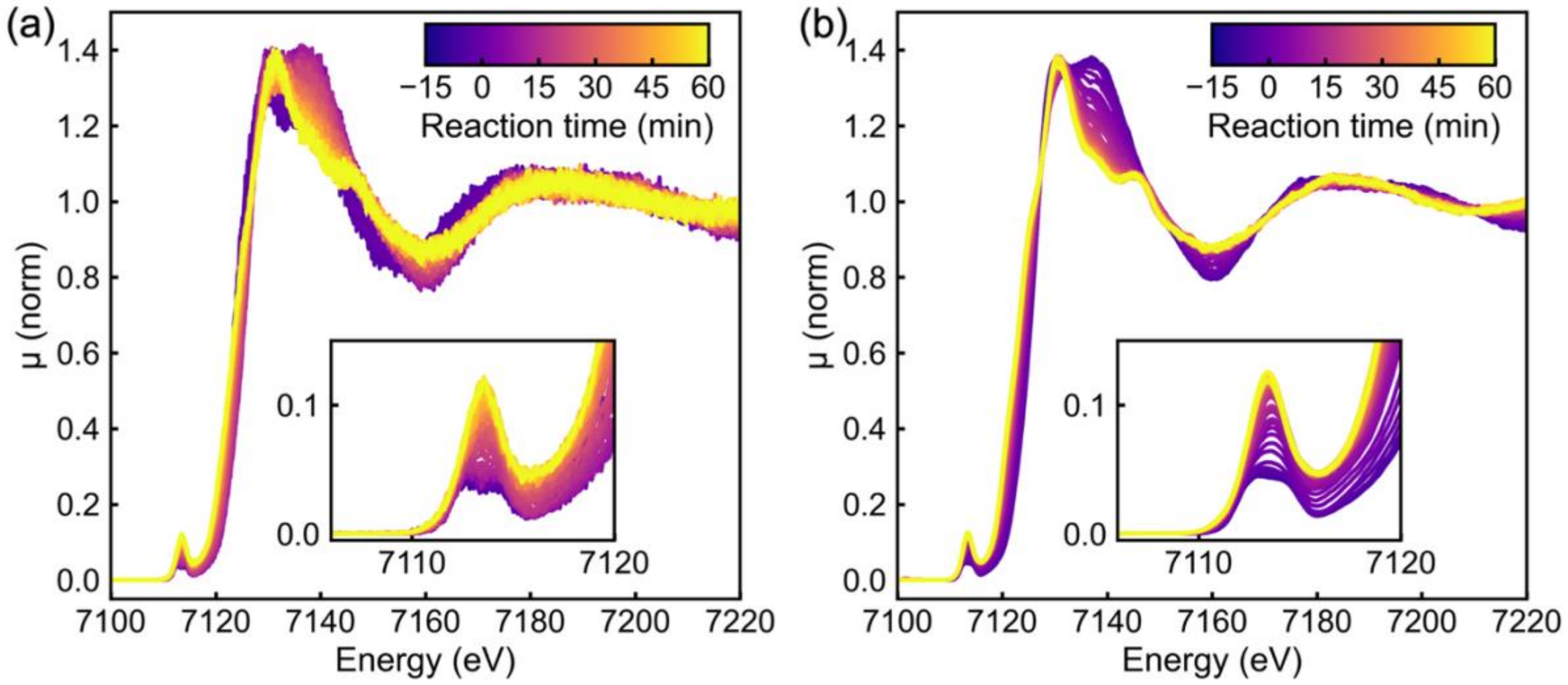

**Figure S11** *In situ* Fe K-edge HERFD-XANES data before (*a*) and after (*b*) processing as discussed in the experimental section.